\documentstyle[aps,multicol,pre,epsf,eqsecnum]{revtex}
\begin{document}
\title
{
Canonical phase space approach to the noisy Burgers equation:\\
Probability distributions
}
\author{Hans C. Fogedby}
\address{
\thanks{Permanent address}
Institute of Physics and Astronomy,
University of Aarhus, DK-8000, Aarhus C, Denmark\\
and\\
NORDITA, Blegdamsvej 17, DK-2100, Copenhagen {\O}, Denmark
}
\maketitle
\begin{abstract}
\noindent
We present a canonical phase space approach to stochastic systems
described by Langevin equations driven by white noise. Mapping the
associated Fokker-Planck equation to a Hamilton-Jacobi equation
in the nonperturbative weak noise limit we invoke a 
{\em principle of least action} for the determination of the
probability distributions. We apply the scheme to the noisy
Burgers and KPZ equations and discuss the time-dependent and stationary
probability distributions. In one dimension we derive the long-time
skew distribution approaching the symmetric stationary Gaussian
distribution. In the short-time region we discuss heuristically the
nonlinear soliton contributions and derive an expression for
the distribution in accordance with the directed polymer-replica
and asymmetric exclusion model results. We also comment on the 
distribution in higher dimensions.
\end{abstract}
\draft
\begin{multicols}{2}
\narrowtext
\section{Introduction}
This is the third of a series of papers where we investigate the noisy Burgers
equation in the context of modelling a growing interface; for a brief account
of the present work we also we refer to \cite{Fogedby98}. In the previous
two papers, in the following denoted paper I \cite{Fogedby98a} 
and paper II \cite{Fogedby98b};
a brief account of paper II also appeared 
in \cite{Fogedby98c}, we discussed the originally proposed one dimensional
noiseless Burgers equation \cite{Burgers29,Burgers74} from a 
solitonic point of view and the noisy one dimensional Burgers equation
\cite{Forster76,Forster77}
in terms of a Martin-Siggia-Rose path integral 
\cite{Martin73,deDominicis75,deDominicis76,Bausch76,Janssen76,deDominicis78},
respectively.

Phenomena far from equilibrium are widespread, including turbulence in fluids,
interface and growth problems, chemical reactions, biological systems,
and even economical and sociological structures. In recent years much
of the focus of modern statistical physics and soft condensed matter has 
shifted towards such systems. 
Drawing
on the case of static and dynamic critical phenomena in and close
to equilibrium \cite{Ma76,Chaikin95}, 
where scaling, critical exponents, and universality
have served to organize our understanding and to provide calculational
tools, a similar approach  has been advanced towards nonequilibrium
phenomena with the purpose of elucidating scaling properties and
more generally the morphology or pattern formation in a driven state.

In this context the noisy Burgers equation provides maybe the simplest
continuum description of an open driven nonlinear system exhibiting 
scaling and pattern formation. The Burgers equation was originally 
suggested
in the one dimensional noiseless version 
\cite{Burgers29,Burgers74,Saffman68,Jackson90,Whitham74}
\begin{equation}
\left(\frac{\partial}{\partial t}-\lambda u\nabla\right)u=
\nu\nabla^2u ~,
\label{bur0}
\end{equation}
as a model for irrotational hydrodynamical fluid flow. Here $u$ is the
irrotational velocity field  and $\nu$ a damping constant or viscosity.
Choosing the nonlinear coupling $\lambda =-1$, we recognize the usual
nonlinear convective term \cite{Landau59a}. 
As a nonlinear model for hydrodynamical
turbulence
\cite{Kida79,Gurbatov81,Aurell92,She92,Episov93,Episov94,Newman97}
Eq. (\ref{bur0}) has been studied intensively. It was
early recognized that the nonlinear structure of the damped velocity
field is dominated by shock waves 
\cite{Burgers74,Whitham74}, yielding an {\em inverse cascade}
and that the Burgers equation does in fact not characterize 
Navier-Stokes 
turbulence which is governed by
a {\em direct cascade} and Kolmogoroff scaling in the inertial regime.

Treating  the field $u$ as the local slope of a growing interface,
we analyzed in paper I, where a more complete bibliography can be found,
the Burgers equation from the point of view
of a soliton-carrying nonlinear damped evolution equation \cite{Scott73}.
The equation describes the transient damped evolution of an initial
slope configuration $u_0$ in terms of a gas of moving {\em right hand}
viscosity-smoothed solitons connected by ramp solutions and with
superposed damped diffusive modes with a gap in their spectrum
proportional to the soliton amplitudes. In a heuristic sense the
transient morphology is thus characterized by two kinds of excitations:
Nonlinear soliton modes and linear gapful diffusive modes.
This picture is also borne out by the nonlinear Cole-Hopf transformation
\cite{Cole51,Hopf50,Halpin95}
\begin{equation}
w(x,t) = \exp{\left[\frac{\lambda}{2\nu}\int^x dx'
~u(x',t)\right]} ~,
\label{ch}
\end{equation}
relating the slope field $u$ to the diffusive field $w$ satisfying the
linear diffusion equation
\begin{equation}
\frac{\partial w}{\partial t} = \nu\nabla^2w ~,
\label{dif}
\end{equation}
with solution
\begin{eqnarray}
w(x,t) &&=\int dx'~ G(x-x',t)w_0(x') ~,
\label{ch1}
\\
G(x,t) &&= [4\pi\nu t]^{-{1\over 2}}\exp{[-x^2/4\nu t]} ~,
\label{ch2}
\\
w_0(x) &&=\exp{\left[\frac{\lambda}{2\nu}\int^x 
dx'~ u_0(x')\right]} ~,
\label{ch3}
\end{eqnarray}
allowing for a steepest descent analysis in the inviscid limit
$\nu\rightarrow 0$ \cite{Aurell92,She92}.

The emphasis on the nonlinear modes in paper I and the insight gained
formed the natural starting point for our study of the noisy Burgers
equation in paper II. This equation has the form 
\cite{Forster76,Forster77,Halpin95,Kardar86,Medina89}
\begin{equation}
\left(\frac{\partial}{\partial t} -\lambda u\nabla\right)u =
\nu\nabla^2 u + \nabla\eta ~,
\label{bur3}
\end{equation}
where the Gaussian white noise driving the equation
is spatially short range correlated according to
\begin{equation}
\langle\eta(x,t)\eta(x',t')\rangle =
\Delta\delta(x-x')\delta(t-t') ~,
\label{noise0}
\end{equation}
characterized by the noise strength $\Delta$.
Equation (\ref{bur3}) has a much richer structure than the 
noiseless counterpart.
The noise is here a singular perturbation in the sense that even a weak
noise strength $\Delta$ eventually drives the morphology described by
Eq. (\ref{bur3}) into a stationary driven state; the characteristic time
scale is set by $t\propto\ln{(1/\Delta)}$ which diverges for
$\Delta\rightarrow 0$. The singular nature of the noise is also
reflected in the known stationary probability distribution for
the slope field \cite{Halpin95,Huse85,Krug92,Krug97}
\begin{equation}
P_{\text{st}}(u)\propto\exp{\left[-\frac{\nu}{\Delta}\int dx~u(x)^2\right]} ~.
\label{dis}
\end{equation}
Here $\Delta$ constitutes an essential singularity in the same
manner as the temperature entering  the usual Boltzmann factor.

Recasting the stochastic Langevin equation (\ref{bur3}) in terms 
of a Martin-Siggia-Rose path integral 
\cite{Martin73,deDominicis75,deDominicis76,Bausch76,deDominicis78,Janssen76}
we proposed 
{\em a principle of least action} in the nonperturbative weak noise
limit $\Delta\rightarrow 0$ and derived
canonical saddle point  or field equations
(note that in paper II the noise field $\varphi$ was {\em rotated},
$\varphi\rightarrow-i\varphi$)
\begin{eqnarray}
\left(\frac{\partial}{\partial t}-\lambda u\nabla\right)u &&=
\nu\nabla^2\varphi ~,
\label{con1}
\\
\left(\frac{\partial}{\partial t}-\lambda u\nabla\right)\varphi &&=
\nu\nabla^2u ~,
\label{con2}
\end{eqnarray}
coupling the slope field $u$ to a deterministic auxiliary 
{\em noise field}
$\varphi$. The coupled field equations (\ref{con1}) and (\ref{con2})
effectively replace the stochastic equation (\ref{bur3}) and describe the
morphology of a growing interface. In addition to the {\em right hand}
soliton already present in the noiseless case as discussed in paper I, the
field equations (\ref{con1}) and (\ref{con2}) also admit an equivalent
{\em left hand} soliton. A growing interface can thus be viewed as a gas of
connected {\em right hand} and {\em left hand} solitons  with superposed
diffusive modes. The weak noise approach also allows a dynamical
description of the soliton and diffusive mode configurations and
associates an action $S$, energy $E$, and momentum $\Pi$ with a particular
interface morphology. The statistical weight of a configuration is 
given by $\exp{[-S/\Delta]}$ with the dynamical action $S$, providing
the generalization of the Boltzmann factor $\exp{[-E/T]}$ for equilibrium
processes to dynamical processes.

The Martin-Siggia-Rose path integral approach moreover permitted
a simple interpretation of the scaling properties of a growing interface.
The dynamical exponent $z=3/2$, entering in the dynamical scaling relation
\cite{Halpin95,Kardar86,Medina89,Krug92,Krug97,Barabasi95,Family85,Jullien85}
\begin{equation}
\langle u(x,t)u(0,0)\rangle =
|x|^{2\zeta-2}F(|t|/|x|^z) ~,
\label{scal}
\end{equation}
is thus related to the gapless soliton dispersion law
\begin{equation}
E\propto \lambda \Pi^z ~,
\label{disp}
\end{equation}
whereas the roughness exponent $\zeta = 1/2$ follows from the spectral
representation
\begin{equation}
\langle u(x,t)u(0,0)\rangle =
\int dK~G(K)\exp{[-iEt+iKx]} ~,
\label{spec}
\end{equation}
assuming that the form factor $G(K)\sim\text{const.}$ in the 
scaling region $K\sim 0$.
The dynamical scaling universality class is associated with the
lowest gapless excitation, i.e., for $\lambda\neq 0$ the soliton mode.
For $\lambda =0$ the Edwards-Wilkinson (EW) universality class 
\cite{Halpin95,Krug97,Edwards82} 
emerges with
a gapless diffusive mode $\omega = \nu k^2$, corresponding to $z=2$,
$\zeta = 1/2$ being unaltered. Furthermore, we derived a heuristic expression
for the scaling function $F$ in terms of the probability distribution
for L\'{e}vy flights \cite{Fogedby94,Fogedby98d};
the scaling function has also been accessed by a mode coupling approach
\cite{Hwa91,Frey96}.

Summarizing, the weak noise saddle point approach to the noisy Burgers
equation advanced in paper II yields a many body description of the
morphology of a growing interface in terms of two kinds of 
{\em quasi-particles}
or {\em elementary excitations}: Nonlinear soliton modes corresponding to the
faceted steplike growth of an interface with superposed linear diffusive
modes. Furthermore, the scaling properties and the notion of universality
classes follow as a byproduct from the dispersion law of the lowest
gapless excitation. For details and references we refer the reader
to the somewhat tutorial presentation in paper II.

Whereas a good understanding of the one dimensional case has been achieved
both by renormalization group methods 
\cite{Forster76,Forster77,Halpin95,Kardar86,Medina89,Barabasi95}
and \cite{Hwa92,Frey94,Tauber95,Frey96}, 
by mapping to directed polymers
\cite{Halpin95}, by mapping to spin chains
\cite{Dahr87,Gwa92a,Gwa92b}, from the lattice exclusion model
\cite{vanBeijeren85,Janssen86,Ligget85}, 
and by the soliton approach in paper II,  the general case 
in $d$ dimensions has proven much more difficult. In $d$ dimensions the
noisy Burgers equation takes the form \cite{Forster76,Forster77}
\begin{equation}
\left(\frac{\partial}{\partial t}-\lambda u_m\nabla_m\right)u_n=
\nu\nabla^2u_n+\nabla_n\eta ~.
\label{bur1}
\end{equation}
Here the longitudinal vector field $u_n$, $n=1,..d$, is associated 
with the height profile $h$ of a growing interface according to
($l_n$ is a line element)
\begin{eqnarray}
u_n&&=\nabla_nh ~,
\nonumber
\\
h(x_n)&&=\int^{x_n}dl_nu_n  ~.
\label{rel}
\end{eqnarray}
Furthermore,
$\nabla^2=\nabla_n\nabla_n$ and we assume summation over repeated
Cartesian indices. The height field $h$  is thus the underlying
{\em potential} for the {\em force} or {\em slope} field $u_n$. 
It follows from
Eqs. (\ref{bur1}) and (\ref{rel}) that $h$ satisfies the
Kardar-Parisi-Zhang (KPZ) equation \cite{Halpin95,Kardar86,Medina89}
\begin{equation}
\frac{\partial h}{\partial t} = \nu\nabla^2 h +
\frac{\lambda}{2}\nabla_nh\nabla_nh
+F+\eta ~,
\label{kpz}
\end{equation}
for a growing interface in $d$ dimensions.
Assuming $\langle\eta\rangle = 0$ we have for completion here
introduced the drift term 
$F=-(\lambda/2)\langle\nabla_nh\nabla_nh\rangle$
in Eq. (\ref{kpz}) in order to ensure that $\langle h\rangle$ decays
in time in a co-moving frame.
In Eqs. (\ref{bur1}) and (\ref{kpz}) $\nu$ is damping constant or
viscosity characterizing the linear diffusive term, $\lambda$ is a
coupling strength for the nonlinear mode coupling or growth term,
and, finally, $\eta$ is a Gaussian white noise driving the equation
and correlated according to
\begin{equation}
\langle\eta(x_n,t)\eta(x_n',t')\rangle =
\Delta\prod_n\delta(x_n-x_n')\delta(t-t') ~,
\label{noise1}
\end{equation}
where $\Delta$ is the noise strength.

In higher dimension dynamic renormalization group calculations
\cite{Halpin95,Kardar86,Medina89} yield a (lower) critical dimension
at $d=2$ and a kinetic phase transition above $d=2$,
separating a smooth phase characterized by the EW universality class
yielding $\zeta=(2-d)/2$ and $z=2$ and  a rough phase characterized by
a strong coupling fixed point. On the transition line renormalization
group calculations and scaling arguments based on the
mapping to directed polymers yield the exponents
$\zeta=0$ and $z=2$ and suggest an upper critical dimension $d=4$
\cite{Laessig95}.
Most recently, an operator expansion method has been applied to 
the strong coupling phase yielding
$(\zeta,z)=(2/5,8/5)$ in $d=2$ and $(\zeta,z)=(2/7,12/7)$
in $d=3$ \cite{Laessig98a,Chin98}.

In the present paper we focus on the stationary and time-dependent
probability distributions for the height and slope fields described
by the Burgers and KPZ  equations (\ref{bur1}) and (\ref{kpz}), respectively.
As discussed in paper II these distributions are basically given by
the Martin-Siggia-Rose path integral weighted by the effective action
for the appropriate paths. In the weak noise limit only the paths governed
by the saddle point equations contribute to the distributions  and, 
as will be discussed here,
we can actually circumvent the path integral formulation 
entirely by a more direct approach based on the Fokker-Planck equation.
In the weak
noise limit this equation takes the form of a Hamilton-Jacobi equation
implying a symplectic structure and immediately lending itself 
to a canonical phase space formulation.
We are thus able in a very direct manner to map the stochastic processes
described by the KPZ-Burgers equations to a conserved dynamical system
with orbits satisfying deterministic canonical Hamilton equations,
identical to the saddle point equations in the path integral approach.
The stochastic nature of the Langevin equations is reflected in
the peculiar topology of the energy surfaces. It turns out that the stationary
probability distribution is determined by an infinite-time orbit on
the zero-energy manifold whereas the time-dependent distribution,
approaching the stationary one at long times, corresponds to a finite-time
orbit. Below we highlight some of our results.

(i) In the generic case of a general nonlinear Langevin equation for
a set of stochastic variables driven
by white noise, the weak noise limit of the associated Fokker-Planck 
equation takes the form of a Hamilton-Jacobi equation, which in
turn implies a symplectic structure with  a {\em principle of least action},  
an action, an 
associated Hamiltonian, 
and 
Hamilton equations of motion. This formulation is equivalent to the
saddle point discussion in the Martin-Siggia-Rose approach in paper II.

(ii) The ensuing canonical phase space formulation allows for a discussion
of the time-dependent probability distributions, i.e., the weak noise
solutions of the Fokker-Planck equation, in terms of phase space orbits
on conserved energy surfaces, governed by  Hamilton equations of motion.
The action associated with an orbit plays the role of a weight function
in much the same way as the Hamiltonian entering the Boltzmann factor
in the description of thermodynamic equilibrium. In the kinetic 
nonequilibrium problem
defined by the Langevin equation the dynamic action yields the probability
distributions.

(iii) In the canonical phase space formulation the underlying stochastic 
nature of the Langevin equation and the relaxational character of the
solutions of the Fokker-Planck equation are reflected in the 
topological structure of the energy surfaces. A structure which differs 
markedly 
from the energy surface topology for ordinary dynamical problems
In particular, the zero-energy
manifold which determines the stationary state, i.e., the stationary probability
distribution, has a two-fold submanifold structure, including a hyperbolic
stationary point, which in the simple case of a single stochastic variable
corresponds to the unstable maximum of an {\em inverted} potential. Moreover,
the {\em waiting time} for the orbits passing close to the stationary point
accounts for the Markovian behavior of the probability distributions.
Finally, the long-time orbit close to the zero-energy manifold
determines via the action the time-dependent probability distribution.

(iv) In the case of a few degrees of freedom the canonical phase space
approach yields the established results following from an analysis of
the Fokker-Planck equations. On the other hand, in the case of 
many degrees of freedom,
i.e., the field theoretical case, the Fokker-Planck equation becomes
an unwieldy multi- dimensional differential-integro equation, and the canonical
phase space approach, replacing the Fokker-Planck equation (in the weak
noise limit) with coupled canonical field equations yields, in addition
to providing an alternative point of view of the stochastic processes in terms
of dynamical system theory, a methodological advantage; particularly in 
the case where we can determine the zero-energy manifold explicitly.

(v) In the field theoretical cases of the noisy Burgers or KPZ equations
in one dimension, we can for special reasons identify the 
zero-energy manifold and determine: (1) the stationary distribution,
(2) the long-time diffusive mode contribution to the time-dependent 
skew distribution and (3) a heuristic expression for the short-time (transient)
soliton mode contribution. In the interesting case of higher dimensions
we are only able to make some general statements.

The paper is organized in the following manner. In Sec II we consider the
generic case of a  nonlinear Langevin equation for many 
stochastic variables driven by white noise. We analyze the
associated
Fokker-Planck equation in the weak noise limit and set up the canonical 
phase space
formulation. In Sec. III we consider as an example
the stochastic overdamped motion in a harmonic potential. In Sec. IV
we apply the formulation to the Burgers and KPZ equations and derive
expressions for the distributions. Finally, in Sec V we present a discussion
and a conclusion.
\section{The canonical phase space approach}
Path integral formulations of the Fokker-Planck equation in the
field theoretical case and aspects of the canonical structure 
have been discussed in the literature, see \cite{Zinn-Justin89} for further
referencing, we believe, however, that the present emphasis on the
canonical phase space formulation as a practical tool is new.
For this purpose we here 
set up the general canonical phase space formalism.
Adhering to the notation in \cite{Zinn-Justin89}
we consider a
general Langevin equation with additive noise \
\cite{Risken89,Fogedby80},
\begin{equation}
\frac{dq_n}{dt} = -\frac{1}{2}F_n(q_m) +\eta_n~.
\label{lan}
\end{equation}
Here $q_n$, $n=1,..N$, is a set of time-dependent stochastic variables.
The index $n$ is discrete 
but is readily generalized later to the field theoretical case of infinitely
many degrees of freedom where $n$ typically includes the spatial variables.
The {\em forces} $F_n(q_m)$ are general functions of $q_n$. In the linear
case of coupled (overdamped) oscillators, $F_n =2\Omega_{nm}q_m(t)$, where 
$\Omega_{nm}$
is a {\em damping matrix}. Finally,
the equation is driven stochastically by a white noise term $\eta_n$
with a Gaussian distribution and correlated 
according to
\begin{equation}
\langle\eta_n(t)\eta_m(t')\rangle = \Delta K_{nm}\delta(t-t') ~.
\label{noise2}
\end{equation}
Here $K_{nm}$ is a constant, symmetric, positive-definite 
{\em noise matrix} of $O(1)$ and  the  correlations are characterized
by  the noise strength 
$\Delta$.

Introducing the notation $\nabla_n=\partial/\partial q_n$ the Fokker-Planck
equation for the (conditional) probability distribution $P(q_n,t,q_n')$
associated with Eq. (\ref{lan}) has the form 
\cite{Zinn-Justin89,Risken89,Fogedby80}
\begin{equation}
\frac{\partial P}{\partial t} = 
\frac{1}{2}\nabla_n[F_nP + \Delta K_{nm}\nabla_m P] ~,
\label{fp1}
\end{equation} 
including a drift term $\nabla_n(F_nP)$, arising from the deterministic
force $F_n$, and a diffusion term $\Delta K_{nm}\nabla_n\nabla_mP$, 
originating from the noise $\eta_n$.

In the equilibrium case, choosing $K_{nm} = \delta_{nm}$ and 
setting $F_n=\nabla_n\Phi$,
corresponding to an effective fluctuation-dissipation theorem and an
underlying thermodynamic free energy $\Phi$, Eq. (\ref{fp1}) admits the 
stationary solution $P_{st}\propto\exp{[-\Phi/\Delta]}$ with $\Delta$
entering as a temperature and mathematically as a singular parameter,
i.e., the Boltzmann distribution.
Using this as a guiding principle we search in the general nonequilibrium
case for
solutions to Eq. (\ref{fp1}) of the form
\begin{equation}
P \propto \exp{\left[-\frac{1}{\Delta}S\right]} ~,
\label{dis2}
\end{equation}  
where the weight function $S$  replaces the free energy $\Phi$ in
the equilibrium case. By
insertion and keeping only terms to leading order in $\Delta$, it is easy
to show that $S$ satisfies an equation of the Hamilton-Jacobi form
\cite{Landau59b,Goldstein80,Arnold89},
\begin{equation}
\frac{\partial S}{\partial t} + H(q_n,\nabla_nS) = 0 ~,
\label{hj}
\end{equation}   
where, introducing the canonical momentum and energy
\begin{eqnarray}
p_n&&=\nabla_nS ~,
\\
E &&= H ~,
\label{mom-e}
\end{eqnarray}    
the conserved energy or Hamiltonian is given by
\begin{equation}
H = \frac{1}{2}K_{nm}p_np_m - \frac{1}{2}F_np_n ~.
\label{ham}
\end{equation}     
From the symplectic structure and dynamical system theory the 
canonical phase space structure follows immediately. The action $S$
has the form \cite{Landau59b}
\begin{equation}
S = \int dt\left(p_n\frac{dq_n}{dt} - H\right) ~,
\label{ac}
\end{equation}
and from the ensuing {\em principle of least action} we derive the Hamiltonian
equations of motion, $dq_n/dt=\partial H/\partial p_n$ and
$dp_n/dt=-\partial H/\partial q_n$,
\begin{eqnarray}
\frac{dq_n}{dt} &&= K_{nm}p_m - \frac{1}{2}F_n ~,
\label{e1}
\\
\frac{dp_n}{dt} &&= \frac{1}{2}p_m\nabla_nF_m ~,
\label{e2}
\end{eqnarray} 
for the orbits in $p_nq_n$ phase space.

The above formulation allows a simple interpretation of the solution
of the Fokker-Planck equation (\ref{fp1}) in the weak noise limit 
$\Delta\rightarrow 0$ in
terms of orbits in a canonical phase space. In order to determine the
transition probability $P(q_n,T,q_n')$ for a configuration $q_n'$
at $t=0$ to a configuration $q_n$ at $t=T$, we simply solve the
Hamilton equations (\ref{e1}) and (\ref{e2}) for an orbit from $q_n'$
to $q_n$ traversed in time $T$ and, subsequently, evaluate the action
according to
Eq. (\ref{ac}), yielding the weight function in Eq. (\ref{dis2}), i.e.,
\begin{equation}
P(q_n,T,q_n')\propto\exp{\left[-\frac{1}{\Delta}\int_{0,q_n'}^{T,q_n}
dt\left(p_n\frac{dq_n}{dt}-H\right)\right]} ~.
\label{dis3}
\end{equation} 
We notice that the relevant orbit is determined by the initial and final
values $q_n'$ and $q_n$ and the elapsed time $T$. The canonically conjugate
momentum $p_n$ is a {\em slaved variable} determined by Eq. (\ref{e2})
and parametrically coupled to Eq. (\ref{e1}). Also, 
unlike the case in ordinary mechanics, the energy $E=H$
in Eq. (\ref{ham}) is not the central quantity in the present interpretation.
The traversal time $T$ is the important variable and the energy manifold
$E(T)$ on which the orbit from $q_n'$ to $q_n$ lies is a function of $T$.

The stochastic nature of the Langevin equation (\ref{lan}) and the
properties of the weak noise solution of the Fokker-Planck equation
(\ref{fp1}) are reflected in the topological submanifold structure of
the energy surfaces in $p_nq_n$ phase space.
Unlike an ordinary mechanical problem $H$ is not bounded from below
and does not separate in a kinetic energy and a potential energy only
depending on $q_n$. In Eq. (\ref{ham}) the potential $-(1/2)F_np_n$
is momentum (velocity) dependent and gives rise to unbounded motion.
Assuming for simplicity that $F_n\rightarrow 0$ for $q_n\rightarrow 0$
the energy surfaces have the submanifold structure depicted in Fig 1.

The origo in phase space constitutes a hyperbolic stationary point,
that is a saddle point determined by the zero-energy submanifold 
$p_n=0$ and the
zero-energy submanifold defined by $K_{nm}p_m-F_n$ being orthogonal to $p_n$,
i.e., $K_{nm}p_m - F_n\perp p_n$.

Assuming $F_n(q_m)\sim 2\Omega_{nm}q_m$ for small $q_n$ the Hamiltonian
(\ref{ham}) is quadratic in $p_n$ and $q_n$ and a stability analysis 
can easily be carried out. In accordance with the present physical
interpretation we assume that the stability  or damping matrix $\Omega_{nm}$
implies an unstable
$p_n=0$ submanifold and a stable submanifold $K_{mn}p_m-F_n \perp p_n$.
The orbits in phase space close to the zero-energy manifold are thus 
those depicted in Fig. 1. In the {\em harmonic oscillator picture} which
applies 
close to the stationary point this behavior corresponds to the motion
in an {\em inverted} parabolic potential as discussed in more detail
in Sec. III.
\begin{figure}
\begin{picture}(100,150)
\put(0.0,-190.0)
{
\centerline
{
\epsfxsize=13cm
\epsfbox{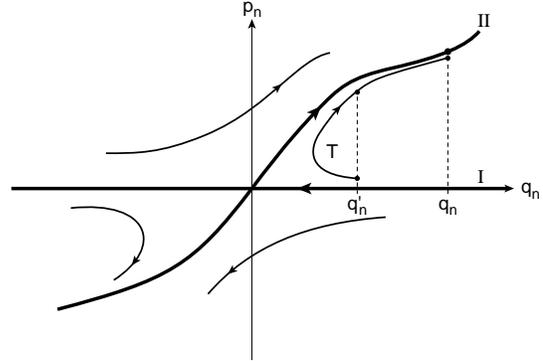}
}
}
\end{picture}
\caption{
Canonical phase space in the general case. The solid curves
indicate the zero-energy {\em transient} submanifold (I) and
{\em stationary} submanifold (II). The stationary saddle point is at the origin.
The finite time ($T$)
orbit from $q'_n$ to $q_n$ migrates to the zero-energy submanifold
for $T\rightarrow\infty$.
}
\end{figure}

The stationary state is given by orbits on the zero-energy manifold
whose structure thus determines the nature of the stochastic problem.
Assuming that $E(T) \propto\exp{[-\text{const.}T]}$ for 
$T\rightarrow\infty$ the stationary state 
\begin{equation}
P_{\text{st}}(q_n)=\lim_{T\rightarrow\infty}P(q_n,T,q_n') ~,
\end{equation}
is obtained from
Eq. (\ref{dis3}), i.e., 
\begin{equation}
P_{\text{st}}(q_n)\propto\exp
{\left[-\frac{1}{\Delta}\int_0^\infty dt p_n\frac{dq_n}{dt}\right]} ~.
\label{dis4}
\end{equation}
We note that for $T\rightarrow\infty$ the orbit from $q_n'$ to $q_n$
converges to the zero-energy manifold, i.e., $P_{\text{st}}$ is determined by
the {\em infinite-time orbits on the zero-energy manifold}.

The basic structure of phase space, depicted in Fig. 1, allows for
a simple dynamical discussion in terms of dynamical system theory
\cite{Jackson90,Arnold89,Ott93} of the approach to the 
stationary state of a damped noise-driven
stochastic system. We first consider an orbit on an $E=0$ surface from
$q_n'$ to $q_n$ in time $T$. The energy surface $E(T)$ depends on $T$
and in the limit $T\rightarrow\infty$, $E\rightarrow 0$ in order
to attain the stationary state. For $E\rightarrow 0$ the initial part
of the orbit moves close to the $p_n = 0$ submanifold and from Eq. (\ref{e1})
is determined by
\begin{equation} 
\frac{dq_n}{dt} = -\frac{1}{2}F_n ~,
\label{detlan}
\end{equation} 
i.e., the deterministic noiseless version of the Langevin equation
(\ref{lan}). In the absence of noise the motion is transient and damped.
Near the {\em transient} submanifold $p_n=0$ the corresponding action
$S\sim 0$ and the probability $P=\text{const.}$, corresponding
to a deterministic behavior. The orbit slows down near the stationary
point  in phase space before it picks up again and moves close
to the other {\em stationary} submanifold $K_{nm}p_m-F_n\perp p_n$. This
part of the orbit carries a finite action $S$, i.e., $P$ depends on $q_n$, 
terminates
in $q_n$ at time $T$, and corresponds for $T\rightarrow\infty$ to the
stationary state. The Markovian
behavior, i.e., the loss of memory  or the independence of the initial
configuration
$q_n'$, is thus associated with the long (infinite) waiting time near (at)
the stationary point.

Whereas the transient submanifold $p_n=0$, yielding $E = 0$,
is consistent with the
Hamiltonian equations (\ref{e1}) and (\ref{e2}) and gives rise to the
deterministic equation (\ref{detlan}), the other possibility of
imposing a zero-energy submanifold by setting $K_{nm}p_m=F_n$ will in
general violate Eqs. (\ref{e1}) and (\ref{e2}). Assuming for
simplicity $K_{nm}=\delta_{nm}$ (note that the symmetric noise
matrix $K_{nm}$ can always be diagonalized by a suitable choice of $p_n$)
we obtain, inserting $p_n=F_n$ in Eqs. (\ref{e1}) and (\ref{e2}), that
the relationship $F_m(\nabla_mF_n-\nabla_nF_m) = 0$ must hold.
In the special case where $F_n=\nabla_n\Phi$, corresponding to an
effective fluctuation-dissipation theorem and an underlying free
energy $\Phi$, the above constraint is trivially satisfied and we obtain the
time-reversed equation of motion $dq_n/dt = (1/2)F_n$ governing the orbit
on the $p_n = F_n$ stationary submanifold. It is interesting to notice 
that the damped motion on the transient $p_n=0$ submanifold is precisely
equal to the time-reversed growing motion on the stationary $p_n=F_n$
submanifold. Finally, it is an easy task to determine the stationary
distribution
\begin{equation}
P_{\text{st}}\propto\exp{\left[-\frac{1}{\Delta}\Phi(q_n)\right]} ~,
\label{stadis}
\end{equation}
by insertion of $p_n=\nabla_n\Phi$ in Eq. (\ref{dis4}) and integrating
over time., in agreement with the solution of the Fokker-Planck equation
(\ref{fp1}) in the stationary case setting $K_{nm}=\delta_{nm}$ and
$F_n=\nabla_n\Phi$.

We are led to the conjecture that for systems in thermal equilibrium
where the force $F_n$ is derived from a thermodynamic free energy $\Phi$,
$F_n=\nabla_n\Phi$, the infinite-time orbits on the stationary
zero-energy submanifold actually converge to the submanifold $p_n=F_n$,
yielding the stationary distribution (\ref{stadis}). In the general
case of a driven stochastic system with a force $F_n$ not derivable
from a free energy the only constraint  is given by
$K_{nm}p_m-F_n \perp p_n$ and we have to solve the Hamilton equations
(\ref{e1}) and (\ref{e2}) in order to determine the phase space orbits.

We finally wish to comment on the connection between the present canonical
phase space approach and the formulation in terms of a Martin-Siggia-Rose
(MSR)
path integral  presented in paper II. For simplicity we consider only the
case of a single stochastic degree of freedom $q$ and set the noise
matrix $K_{nm}=1$.

All the relevant properties are extracted from the generator
\cite{Zinn-Justin89}
\begin{equation}
Z(\mu) = \langle\exp{[i\int dt~\mu(t)q(t)]}\rangle ~,
\label{gen}
\end{equation}
where $\langle\cdots\rangle$ denotes an average with respect to the 
Gaussian noise distribution
\begin{equation}
P(\eta)\propto \exp{\left[-\frac{1}{2\Delta}\int dt~\eta(t)^2\right]} ~.
\label{ndis}
\end{equation}
The Langevin equation (\ref{lan}) for one degree of freedom is enforced
by the delta function constraint
$\int\prod_tdtJ\delta(dq/dt+(1/2)F-\eta)=1$, where
the Jacobian $J=\exp{[(1/4)\int dtdF/dq]}$ \cite{Zinn-Justin89}.
Exponentiating the constraint
and in the process introducing an additional {\em noise variable} 
$p$, averaging over
the noise $\eta$ according to Eq. (\ref{ndis}), and scaling $p$,
$p\rightarrow p/\Delta$, we obtain the expression
\begin{equation}
Z(\mu)\propto\int\prod_tdpdq
\exp{\left[\frac{i}{\Delta}S_{\text{MSR}}\right]}
\exp{\left[i\int dt~\mu(t)q(t)\right]} ~,
\label{gen2}
\end{equation}
where the action has the Feynman form \cite{Das93,Feynman65}
\begin{equation}
S_{\text{MSR}} = \int dt\left(p\frac{dq}{dt} - H_{\text{MSR}}\right) ~,
\label{msra}
\end{equation}
with complex Hamiltonian
\begin{equation}
H_{\text{MSR}} = 
-\frac{1}{2}pF - \frac{i}{2}p^2 + i\frac{\Delta}{4}\frac{dF}{dq} ~.
\label{msrh}
\end{equation}
For the probability distribution $P(q,t,q')$ we find in particular
\begin{equation}
P(q,t,q')\propto
\int^q_{q'}
\prod_tdpdq\exp{\left[\frac{i}{\Delta}S_{\text{MSR}}\right]} ~,
\label{probdis}
\end{equation}
where $S_{\text{MSR}} = \int_0^tdt(pdq/dt - H_{\text{MSR}})$ 
and the path integral samples
all orbits from $q'=q(0)$ to $q=q(t)$ weighted with $S_{\text{MRS}}$; 
note that
the noise field ranges freely.

Reconstructing the underlying {\em quantum mechanics}, yielding the
path integral (\ref{probdis}), $P(q,t,q')$ can be regarded as a matrix
element of the evolution operator $\exp{[-i\hat{H}_{\text{MSR}}t]}$ in
a q-basis,
$
P(q,t,q')\propto\langle q|\exp{(-i\hat{H}_{\text{MSR}}t)}|q'\rangle
$
,
where $\hat{H}_{\text{MSR}}$ is the {\em quantum} version of 
$H_{\text{MSR}}$.
It follows that  $P(q,t,q')$ then satisfies the {\em Schr\"{o}dinger
equation}
\begin{equation}
i\Delta\frac{\partial P}{\partial t} = \hat{H}_{\text{MSR}}P ~,
\label{se}
\end{equation}
with $\Delta$ playing the role of an effective {\em Planck constant}.
The noise variable $p$ becomes the momentum operator
$\hat{p}=-i\Delta d/dq$ and we obtain, inserting $p\rightarrow\hat{p}$
in Eq. (\ref{msrh}),
\begin{equation}
\hat{H}_{\text{MSR}} = i\frac{1}{2}
\left[\Delta^2\frac{d^2}{dq^2}+ i(\hat{p}F)_{\text{order}}+ \frac{\Delta}{2}
\frac{dF}{dq}\right] ~,
\label{qham}
\end{equation}
where since $[\hat{p},q]=-i\Delta$ and $F$ depends on $q$ we still have 
to specify the ordering in $(\hat{p}F)_{\text{order}}$.

Comparing Eqs. (\ref{se}) and (\ref{qham}) with the Fokker-Planck
equation (\ref{fp1}) in the present case,
\begin{equation}
\frac{\partial P}{\partial t} = \frac{1}{2}\left[\Delta\frac{d^2}{dq^2} 
+ F\frac{d}{dq} + \frac{dF}{dq}\right]P ~,
\label{fopl}
\end{equation}
we find agreement provided we choose the symmetric ordering
$
(\hat{p}F)_{\text{order}}=\frac{1}{2}(\hat{p}F + F\hat{p})
$
.
Alternatively, we are free to choose a {\em normal ordering}
$
(\hat{p}F)_{\text{order}}=\hat{p}F
$
and neglect the Jacobian contribution $(\Delta/2)dF/dq$ in Eq. (\ref{qham}).
The Fokker-Planck equation then becomes the underlying
{\em Schr\"{o}dinger
equation} for the path integral with a complex non Hermitian Hamiltonian,
\begin{equation}
\hat{H}_{\text{FP}} = 
i\frac{1}{2}\left[\Delta^2\frac{d^2}{dq^2}+ \Delta\frac{d}{dq}F\right] ~.
\label{qham1}
\end{equation}
The non Hermitian form of $\hat{H}_{\text{FP}}$ with the $\hat{p}$ operator
on the left ensures that Eq. (\ref{fopl}) has the form of a conservation
law ensuring the conservation of probability.

In the limiting case $F=2\omega q$ the Hamiltonian $H_{\text{MSR}}$ 
describes
a harmonic oscillator and it is easy to see that the Jacobian contribution
$i(\Delta/4)dF/dq=i(\Delta/2)\omega$ in $\hat{H}_{\text{MSR}}$ 
precisely cancels
the zero point motion and ensures a stationary state for $t\rightarrow\infty$.

Finally, in the {\em classical}
weak noise limit for $\Delta\rightarrow 0$, the path integral 
(\ref{probdis}) is dominated by
the {\em classical} or stationary orbits following from the {\em principle
of least action} $\delta S_{\text{MSR}} = 0$ and determined as solutions of
the
Hamilton equations of motion: 
$dq/dt=\partial H_{\text{MSR}}/\partial p$ and
$dp/dt=-\partial H_{\text{MSR}}/\partial q$; the distribution being given by
$P\propto\exp{[(i/\Delta)S_{\text{MSR}}]}$.
This procedure is, however, entirely equivalent to performing the limit
$\Delta\rightarrow 0$ directly in the Fokker-Planck equation, yielding
the Hamilton-Jacobi equation (\ref{hj}) and the present canonical phase space
approach.
\section{The harmonic oscillator - an example}
In order to illustrate the phase space method we here apply it
to the simple
case of an overdamped harmonic oscillator described by the Langevin
equation and noise correlations
\begin{eqnarray}
&&\frac{dq}{dt} = -\omega q + \eta ~,
\label{lan6}
\\
&&\langle\eta(t)\eta(t')\rangle = \Delta\delta(t-t') ~,
\label{noise3}
\end{eqnarray}
with associated Fokker-Planck equation
\begin{equation}
\frac{\partial P}{\partial t} = \frac{1}{2}
\frac{\partial}{\partial q}
\left[\Delta\frac{\partial P}{\partial q} + 2\omega P\right] ~.
\label{fp2}
\end{equation}
This system is well-known and easily analyzed 
\cite{Zinn-Justin89,Risken89}.
The time-dependent
probability distribution $P(q,T,q')$, the solution of Eq. (\ref{fp2}),
is given by
\begin{equation} 
P(q,T,q')\propto 
\exp{\left[-\frac{\omega}{\Delta}
\frac{[q-q'\exp{(-\omega T)}]^2}
{1-\exp{(-2\omega T)}}\right]} ~,
\label{dis5}
\end{equation}
approaching the stationary distribution
\begin{equation} 
P_{\text{st}}(q)\propto\exp{\left[-\frac{1}{\Delta}\omega q^2\right]} ~,
\label{stadis2}
\end{equation}
in the limit $T\rightarrow\infty$.

We now proceed to derive these results within the canonical phase
space formulation. Since $F=2\omega q$ and $K=1$ for one degree of
freedom, we obtain from Eqs. (\ref{ham}) - (\ref{e2})
the Hamiltonian
\begin{equation}
H = \frac{1}{2}p^2 - \omega qp ~,
\label{ham2}
\end{equation}
the action
\begin{equation}
S = \int dt\left(p\frac{dq}{dt}-H\right) ~,
\label{ac2}
\end{equation}
and the Hamilton equations of motion
\begin{eqnarray}
\frac{dq}{dt} &&= p-\omega q ~,
\label{eq1}
\\
\frac{dp}{dt} &&= \omega p ~.
\label{eq2}
\end{eqnarray}
The phase space is depicted in Fig. 2 and corresponds to the
vicinity of the stationary point in Fig. 1 for one degree of freedom.

For a single degree of freedom we can explicitly determine both
zero-energy submanifolds: $p=0$ and $p=2\omega q$, and determine the
orbits. On the transient $p=0$ submanifold $dq/dt=-\omega q$, i.e.,
the deterministic equation of motion for $\eta = 0$, with a damped
solution $q=q_0\exp{[-\omega t]}$, $q_0=q(0)$, corresponding to
the damped orbit approaching the stationary saddle point at 
$(p,q)=(0,0)$. The action associated with this orbit is $S=0$,
i.e., $P=\text{const.}$, characterizing the deterministic motion. 
On the stationary 
submanifold $p=2\omega q$ the Hamiltonian equations coincide,
$dq/dt=2\omega q$, and we obtain a growing solution $q=q_0\exp{[\omega t]}$,
$q_0=q(0)$, associated with the orbit emerging from the stationary point.
As discussed in Sec. II we note that the {\em stationary} orbit is the
time-reversed {\em mirror} of the {\em transient} orbit.

The complete solution of Eqs. (\ref{eq1}) and (\ref{eq2}) is also
easily obtained. For an orbit from $q'$ to $q''$ in time $T$ and, noting
that $p$ is a slaved variable, we obtain
\begin{eqnarray}
q(t) &&=
\frac{q''\sinh{\omega t}+q'\sinh{\omega(T-t)}}{\sinh{\omega T}} ~,
\label{sol1}
\\
p(t) &&=
\omega\frac{q''e^{\omega t}-q'e^{-\omega(t-T)}}
{\sinh{\omega T}} ~.
\label{sol2}
\end{eqnarray}
\begin{figure}
\begin{picture}(100,150)
\put(0.0,-190.0)
{
\centerline
{
\epsfxsize=13cm
\epsfbox{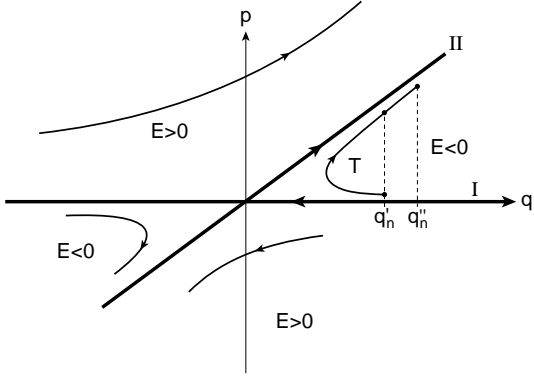}
}
}
\end{picture}
\caption{
Canonical phase space plot in the case of an overdamped oscillator.
The solid curves
indicate the zero-energy {\em transient} submanifold (I) and
{\em stationary} submanifold (II). The stationary saddle point is at the origin.
The finite time ($T$)
orbit from $q'_n$ to $q_n$ migrates to the zero-energy submanifold
for $T\rightarrow\infty$. We have also indicted the sign of the
energy $E$ in the four domains.
}
\end{figure}
For large $T$ the noise variable $p$ is initially close to zero,
corresponding  to the transient deterministic regime; for $t$ close to
$T$ $p$ eventually  leaves zero and approaches the limiting value $2\omega q''$,
corresponding to the stationary regime. Likewise, we note that $q\sim 0$
for $T\rightarrow\infty$ for most $t$. The behavior of $q$ and $p$ is
depicted in Fig. 3.
\begin{figure}
\begin{picture}(100,150)
\put(0.0,-190.0)
{
\centerline
{
\epsfxsize=13cm
\epsfbox{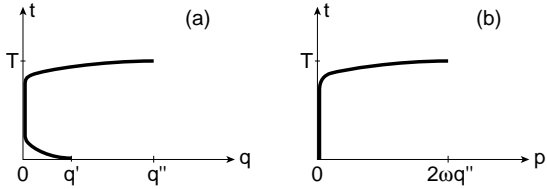}
}
}
\end{picture}
\caption{
Plot of $q$ and $p$ as functions of $t$ in the case of an 
overdamped oscillator.
In a) we depict the dependence of $q$; for large $T$ the coordinate
$q$ stays close to the stationary saddle point. In b) we show the dependence
of $p$; for large $T$ the momentum $p$ is initially  close to the
{\em transient} submanifold $p=0$ but eventually moves on to the
{\em stationary} submanifold $p=2\omega q$.
}
\end{figure}

The orbit from $q'$ to $q''$ traversed in time $T$ lies on the energy surface
given by
\begin{equation}
E=\frac{\omega^2}{2}
\frac{q''^2+q'^2-2q'q''\cosh{\omega T}}{\sinh^2{\omega T}} ~.
\label{en}
\end{equation}
For fixed $q'$ and $q''$ the energy $E$ is a function of $T$, $E = E(T)$.
In the limit $T\rightarrow\infty$, $E\rightarrow 0$ and the orbit converges
to the zero-energy manifold as indicated in Fig. 2. The actual $T=\infty$
orbit passing through the stationary point is then defined by the
limiting orbit for $T\rightarrow\infty$. 

Moreover, from the Hamilton equations (\ref{eq1}) and (\ref{eq2}) 
we readily deduce the 
second order {\em Newton equation of motion}
\begin{equation} 
\frac{d^2q}{dt^2} = \omega^2 q ~,
\label{ne}
\end{equation}
describing the orbit in an inverted harmonic potential
$-(1/2)\omega^2q^2$, and allowing for a simple discussion of the motion
in $pq$ phase space.

The finite energy orbits fall in two categories depending on the sign of $E$.
For $E>0$ the orbits pass through the unstable maximum of the inverted
potential with finite momentum; for $E<0$ the unbounded orbits are confined
by the potential to either positive or negative values of $q$. The limiting
case $E=0$ corresponds to an orbit approaching the maximum, the hyperbolic
stationary point, with zero momentum. This point represents an unstable
equilibrium where the {\em particle} spends an infinite amount of time,
corresponding to the establishment of Markovian behavior, i.e.,
the loss of memory and independence of the initial configuration $q''$. 
The motion is
depicted in Fig. 4.
\begin{figure}
\begin{picture}(100,150)
\put(0.0,-190.0)
{
\centerline
{
\epsfxsize=13cm
\epsfbox{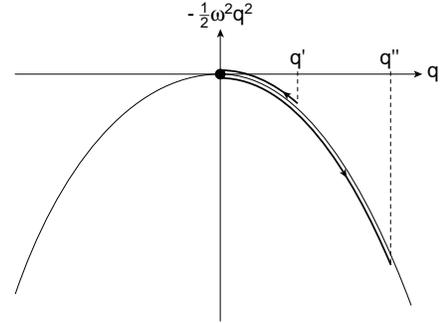}
}
}
\end{picture}
\caption{
In the case of the overdamped oscillator the orbits in $pq$
phase space depicted in Fig. 2 corresponds in $q$ space to
the motion in an {\em inverted} parabolic potential. The unstable
maximum corresponds to the stationary saddle point.
}
\end{figure}

In terms of the explicit solution (\ref{sol1}) and (\ref{sol2}) we finally 
derive the action associated with the orbit,
\begin{equation} 
S = \omega\frac{(q''-q'e^{-\omega T})^2}{1-e^{-2\omega T}} ~,
\label{ac3}
\end{equation} 
and recover from $P\propto\exp{[-S/\Delta]}$ the time-dependent and in 
the limit $T\rightarrow\infty$, stationary distributions (\ref{dis5}) and
(\ref{stadis2}), respectively.

The Hamiltonian (\ref{ham2}) and the equations of motion (\ref{eq1})
and (\ref{eq2}), yielding the canonical phase space representation of the
Langevin equation for a noise-driven overdamped harmonic oscillator, have
the same structure as the dynamical description of an ordinary harmonic
oscillator. Shifting the momentum, $p\rightarrow p+\omega q$, 
$H\rightarrow(1/2)p^2-(1/2)\omega q^2$, describing the motion in an 
inverted harmonic potential as discussed above. The equations
of motion now take the form $dq/dt=p$ and $dp/dt=\omega q$ with solutions
given as linear combinations of a growing and a damped solution as in
Eqs. (\ref{sol1}) and  (\ref{sol2}). However, performing a Wick rotation 
$t\rightarrow it=\tau$ in combination with the transformation to a complex
momentum $p\rightarrow -ip$, note that $p$ is basically a dummy variable
representing the noise, we obtain $H\rightarrow -H_{\text{osc}}$, where 
$H_{\text{osc}}$
is the oscillator Hamiltonian
$H_{\text{osc}} = \frac{1}{2}p^2 + \frac{1}{2}\omega^2 q^2$, 
yielding the equations of motion $dq/d\tau=p$, and $dp/d\tau=-\omega q$
with bounded periodic motion in imaginary time $\tau$. The energy 
$E=H_{\text{osc}}$
is positive and the finite energy orbits in $pq$ phase space move on 
concentric ellipses. The zero-energy manifold corresponds to the 
origin $(p,q)=(0,0)$ in phase space. In fact, subject to the Wick rotation 
$t\rightarrow it$ and a {\em rotation} of $p$, $p\rightarrow -ip$, this phase
space structure is mapped to the phase space structure in Fig. 2 with
a hyperbolic stationary point and unbounded orbits.
As pointed out in Sec. II it is precisely in the energy surface topology
that the stochastic problem differs from an ordinary dynamical problem;
here exemplified in the context of the oscillator.
Correspondingly, the action (\ref{ac2}) for an orbit from $q'$ to $q''$
in time $iT$ transforms to 
$-iS_{\text{osc}}+(\omega/2)(q''^2-q'^2)$ where $S_{\text{osc}}$
is the action for a harmonic oscillator \cite{Das93,Feynman65}
\begin{equation}
S_{\text{osc}} =
\frac{\omega}{2\sin{\omega T}}
[(q''^2+q'^2)\cos{\omega T}-2q'q''] ~,
\label{ac4}
\end{equation} 
yielding $P(q'',iT,q')$ in accordance with with Eq. (\ref{dis5}).

Finally, we present a simple calculation of the leading correction at
long times to the stationary distribution (\ref{stadis2}) which will
prove useful in our discussion of the Burgers-KPZ equations in the
next section. For $T\rightarrow\infty$ the orbit in phase space is close
to the stationary zero-energy submanifold $p=2\omega q$. Replacing the orbit
from $q'$ to $q''$ at long times $T$ with an orbit {\em on the stationary
manifold} we obtain a constraint which allows us to simply
evaluate the correction to $P_{\text{st}}$. Consequently, inserting the
zero-energy constraint
$p=2\omega q$ in the canonical equation (\ref{eq1}) yields
$dq/dt=\omega q$ with solution $q''=\exp{(\omega T)}q'$ for $q'$ and $q''$
on the $p=2\omega q$ manifold. Inserting this solution in the action
(\ref{ac2}) we obtain
\begin{equation}
S=\omega(q''^2-q'^2)=\omega q''^2(1-\exp{(-2\omega T)}) ~,
\label{ac5}
\end{equation}  
in accordance with an expansion of the exact result (\ref{ac3}) to 
leading order in $\exp{(-\omega T)}$.
\section{Canonical formulation of the Burgers - KPZ equations}
In this section we apply the canonical phase space method developed in
Sec. II to the Burgers and KPZ equations
(\ref{bur1}) and (\ref{kpz}).
\subsection{The general case}
First applying the canonical formulation to the Burgers equation (\ref{bur1})
the index $n$ in Sec. II now comprises both the continuous spatial
coordinate $x_n$, $n=1,..d$ and the vector index of the slope field $u_n$,
$n=1,..d$, i.e., $n\rightarrow x_n,n$. Furthermore, we choose the noise
matrix $K_{nm}$ and the forces $F_n$ according to the prescription,
$K_{nm}\rightarrow\nabla^2\prod_n\delta(x_n-x_n')$ and
$F_n\rightarrow -2(\nu\nabla^2u_n+\lambda u_m\nabla_mu_n)$.
With the identification $q_n\rightarrow u_n(x_m)$ and 
$p_n\rightarrow p_n(x_m)$ we thus obtain the Burgers action and  Hamiltonian
density
\begin{eqnarray} 
S_{\text{B}} &&= \int_0^Td^dxdt
\left(p_n\frac{\partial u_n}{\partial t} - {\cal H}_{\text{B}}\right) ~,
\label{ac6}
\\
{\cal H}_{\text{B}} &&= p_n\left(\nu\nabla^2u_n + \lambda u_m\nabla_mu_n - 
\frac{1}{2}\nabla_n\nabla_mp_m\right) ~,
\label{ham4}
\end{eqnarray}
and the ensuing Hamilton equations of motion
\begin{eqnarray}
\left(\frac{\partial}{\partial t} - \lambda u_m\nabla_m\right)u_n
=&& \nu\nabla^2u_n-\nabla_n\nabla_mp_m ~,
\label{b1}
\\
\left(\frac{\partial}{\partial t} - \lambda u_m\nabla_m\right)p_n
=&& -\nu\nabla^2p_n
\nonumber
\\
 && + \lambda(p_n\nabla_mu_m-p_m\nabla_nu_m) ~.
\label{b2}
\end{eqnarray}
The time-dependent probability distribution is then in the weak noise
limit given by
\begin{equation}
P(u_n,T,u_n')\propto
\exp{\left[-\frac{1}{\Delta}S_{\text{B}}(u_n,T,u_n')\right]} ~.
\label{dis6}
\end{equation}
The Hamilton equations (\ref{b1}) and (\ref{b2}) determining the orbits
in $p_nu_n$ phase space thus replace the noisy Burgers equation (\ref{bur1})
in the weak noise limit and the distribution (\ref{dis6}), evaluated for
an appropriate orbit from $u_n'$ to $u_n$ traversed in time $T$, constitutes a
weak noise solution of the Fokker-Planck equation associated with the
Burgers equation,
\begin{eqnarray}
&&\frac{\partial P(u_n,t)}{\partial t} = 
\nonumber
\\
&&- \int d^dx\frac{\delta}{\delta u_n}
\left[(\nu\nabla^2u_n+\lambda u_m\nabla_mu_n)P(u_n,t)\right]
\nonumber
\\
&&+\frac{\Delta}{2}\int d^dxd^dx'\frac{\delta^2}{\delta u_n\delta u_n'}
\left[\nabla^2\prod_n\delta(x_n-x_x')P(u_n,t)\right] ~.
\nonumber
\\
\label{fp3}
\end{eqnarray}
The time-dependent and stationary distributions are determined by the
orbits near and on the zero-energy manifold. 
From the general discussion in Sec. II it follows that 
the zero-energy manifold has a submanifold structure with
a transient $p_n=0$ submanifold, a hyperbolic stationary point at
$(u_n,p_n) = (0,0)$, and a stationary submanifold defined by 
$\nu\nabla^2u_n+\lambda u_m\nabla_mu_n - (1/2)\nabla_n\nabla_mp_m$
{\em orthogonal} to $p_n$; here treating the integration over $x$ in 
$\int d^dx{\cal H}_B$ as an {\em inner product}.
On the transient submanifold $p_n=0$ Eq. (\ref{b2}) is trivially satisfied 
and the orbits are governed by 
the noiseless Burgers equation 
\begin{equation}
\left(\frac{\partial}{\partial t} - \lambda u_m\nabla_m\right)u_n 
= \nu\nabla^2u_n
\label{bur4}
\end{equation}
which is analyzed by means of the Cole-Hopf transformation  (\ref{ch})
$u_n=\nabla_n h$, $h = (2\nu/\lambda)\ln{w}$, with $w$ satisfying
Eq. (\ref{dif}) and solved by means of the Green's function (\ref{ch2})
generalized to the $d$-dimensional case. On the other hand, on the stationary
submanifold defined by $\int d^dx {\cal H}_{\text{B}} = 0$, determining the
stationary distribution $P_{\text{st}}(u_n)$, the orbits are given by the
coupled equations (\ref{b1}) and (\ref{b2}) and will be discussed in the
next section.

For later reference we also present the canonical formulation of the
KPZ equation (\ref{kpz}). Here we choose $q_n\rightarrow h(x)$,
$p_n\rightarrow p(x)$, $K_{nm}\rightarrow\prod_n\delta(x_n-x_n')$,
and $F_n\rightarrow -2(\nu\nabla^2h+(\lambda/2)
\nabla_nh\nabla_nh)$,
and obtain action, Hamiltonian density, and equations of motion,
\begin{eqnarray}
S_{\text{KPZ}} &&= 
\int^T_0 d^dxdt\left(p\frac{\partial h}{\partial t}-{\cal H}_{\text{KPZ}}
\right) ~,
\label{ac7}
\\
{\cal H}_{\text{KPZ}} &&= p\left(\nu\nabla^2h+
\frac{\lambda}{2}\nabla_nh\nabla_nh
+ \frac{1}{2}p\right) ~,
\label{ham5}
\\
\frac{\partial h}{\partial t} &&= \nu\nabla^2h+\frac{\lambda}{2}
\nabla_nh\nabla_nh + p ~,
\label{kpz1}
\\
\frac{\partial p}{\partial t} &&= -\nu\nabla^2p+\lambda\nabla_n(p\nabla_nh) ~,
\label{kpz2}
\end{eqnarray}
yielding the weak noise distribution
\begin{equation}
P(h,T,h') \propto \exp{\left[-\frac{1}{\Delta}S_{\text{KPZ}}(h,T,h')\right]} ~,
\label{dis7}
\end{equation}
as solution of the Fokker - Planck equation
\begin{eqnarray}
&&\frac{\partial P(h,t)}{\partial t} 
=
\nonumber
\\
&& -\int d^dx\frac{\delta}{\delta h}
\left[(\nu\nabla^2h+\frac{\lambda}{2}\nabla_nh\nabla_nh)P(h,t)\right]
\nonumber
\\
&& + \frac{\Delta}{2}\int d^dxd^dx'
\frac{\delta^2}{\delta h\delta h'}
\left[\prod_n\delta(x_n-x_n')P(h,t)\right] ~.
\label{fp4}
\end{eqnarray}

The KPZ formulation is, however, completely equivalent to the Burgers
formulation. In Eq. (\ref{b1}) only the longitudinal component of
the noise field $p_n$ couples to the longitudinal slope field
$u_n = \nabla_nh$ and we can without loss of generality assume that
$p_n=\nabla_n\phi$ is purely longitudinal since Eq. (\ref{b2}) is
linear in $p_n$; this property reflects the conserved noise $\nabla_n\eta$
driving the Burgers equation (\ref{bur1}). Comparing Eqs (\ref{b1})
and (\ref{b2}) with Eqs. (\ref{kpz1}) and (\ref{kpz2}) we obtain
complete equivalence by choosing $\nabla_mp_m=-p$ or $p=-\nabla^2\phi$.
\subsection{The one dimensional case}
In one dimension and focussing on the slope field $u$, which in many
respects is the {\em natural} variable in discussing a growing interface,
the canonical equations (\ref{b1}) and (\ref{b2}) take the simple 
form
\begin{eqnarray}
\left(\frac{\partial}{\partial t} - \lambda u\nabla\right)u 
&&= \nu\nabla^2 u - \nabla^2 p ~,
\label{b11}
\\
\left(\frac{\partial}{\partial t} - \lambda u\nabla\right)p
&&= -\nu\nabla^2 p ~,
\label{b12}
\end{eqnarray}
originating from the Hamiltonian density
\begin{eqnarray}
{\cal H}_{\text{B}} = 
p\left(\nu\nabla^2u+\lambda u\nabla u - \frac{1}{2}\nabla^2 p\right) ~.
\label{hamb1}
\end{eqnarray}
We note that both $u$ and $p$ are scalar fields and that the
$\lambda$-dependent term on the RHS of (\ref{b2}) cancels.
Also, subject to the shift transformation $p = \nu(u-\varphi)$ Eqs. (\ref{b11})
and (\ref{b12}) are identical to the equations (\ref{con1}) and (\ref{con2})
discussed in paper II.

It is an important property of the one dimensional case that we can 
determine the explicit form of the stationary zero-energy
manifold, as was the trivial case for one degree of freedom discussed
in Sec. II.
For
\begin{equation}
p = 2\nu u ~,
\label{sub}
\end{equation}
the canonical equations (\ref{b11}) and (\ref{b12}) become identical
and the energy density (\ref{hamb1}) takes the form of a total derivative,
${\cal H}_{\text{B}}\rightarrow(2/3)\lambda\nu\nabla u^3$, yielding a vanishing
total energy $E_{\text{B}}=\int {\cal H}_{\text{B}} = 0$ 
for vanishing slope field at the
boundaries. Owing to the vector character of $u_n$ and $p_n$ and the
presence of the $\lambda(p_n\nabla_mu_m-p_m\nabla_nu_m)$ term 
in Eq. (\ref{b2}) such a 
transformation does not seem possible for $d>1$ and we dont have a
similar {\em contraction} of the stationary submanifold.

In other words, in $d=1$ the orbit from $u'$ to $u''$ in time
$T$ for $T\rightarrow\infty$ does not only approach the zero-energy
submanifolds $p=0$ and $\nu\nabla^2u+\lambda u\nabla u-\frac{1}{2}\nabla^2p$
orthogonal to $p$ but actually converges to the submanifold $p=2\nu u$
on the stationary submanifold. This phase space  behavior is depicted
in Fig. 5. In Fig. 5a we show the {\em contraction} of the stationary
manifold. In Fig. 5b we depict the orbits in $pu$ phase space in a 
similar manner as in Fig. 2.

We finally wish to present a plausibility  argument for the attraction of 
the orbits to the submanifold given by Eq. (\ref{sub}). Denoting the
deviation from the submanifold by $\delta u$ and inserting 
$p=2\nu(u+\delta u)$ in Eqs. (\ref{b11}) and (\ref{b12}) we obtain
to leading order in $\delta u$,
\begin{equation}
\left(\frac{\partial}{\partial t} - \lambda u\nabla\right)\delta u
=\nu\nabla^2\delta u ~.
\label{dev}
\end{equation} 
Noting that $\partial/\partial t - \lambda u\nabla$ in invariant
under the Galilean transformation $x\rightarrow x-\lambda u_0 t$,
$u\rightarrow u+ u_0$ and choosing a local frame with vanishing $u$,
the Fourier modes 
$\delta u_k\propto\exp{[-\nu k^2t]}\rightarrow 0$ for large $t$, 
implying that
the orbits approach the zero-energy submanifold.

\begin{figure}
\begin{picture}(100,150)
\put(0.0,-190.0)
{
\centerline
{
\epsfxsize=13cm
\epsfbox{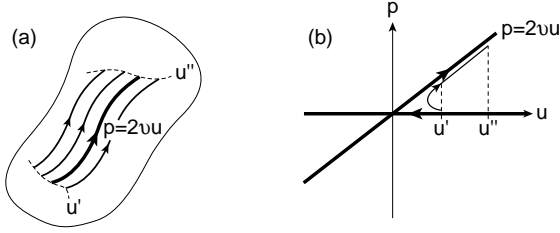}
}
}
\end{picture}
\caption{
Here we depict the phase space behavior in the case of the
noisy one dimensional Burgers equation.
In a) we show the 
{\em contraction} to the zero-energy submanifold $p=2\nu u$,
characteristic of the one dimensional case. In b) we show similar to
Fig. 2 the orbits in $pu$ phase space.
}
\end{figure}
\section{Discussion and conclusion}
In this final section we derive results for the probability distributions
and attempt to draw some general conclusions on the basis of the 
canonical phase space approach to the
noisy Burgers and KPZ equations presented in the previous sections.
\subsection{The one dimensional case}
The time-dependent distribution (\ref{dis6}) is determined by the form
of the action $S_{\text{B}}(u,T)$ in Eq. (\ref{ac6}). The following analysis
implies that the action has the generic form 
\begin{equation}
S_{\text B} = S_{\text{st}}(u) + S_{\text{diff}}(u,T) + S_{\text{sol}}(u,T) ~,
\label{genform}
\end{equation}
where $S_{\text{st}}$ yields the stationary distribution (\ref{dis}), 
$S_{\text{diff}}$ gives rise to corrections due to the linear diffusive
modes, and
$S_{\text{sol}}$ originates from the nonlinear soliton modes;
both $S_{\text{diff}}$ and
$S_{\text{sol}}$ must vanish in the limit $T\rightarrow\infty$ so that we
attain the stationary distribution given by $S_{\text{st}}$.

In the linear Edwards-Wilkinson case for $\lambda = 0$ only 
diffusive modes contribute and there 
is no growth. In wave number space the field equations (\ref{b11}) and
(\ref{b12}) take the same form as Eqs. (\ref{eq1}) and (\ref{eq2}) in Sec. III.
A straightforward generalization of Eqs. (\ref{sol1}) and (\ref{ac3}) then
yields the orbit, $u''_k = u_k(T)$, $u'_k = u_k(0)$,  and $\omega_k = \nu k^2$,
\begin{equation}
u_k(t)=
\frac{u''_k\sinh\omega_kt+u'_k\sinh\omega_k(T-t)}{\sinh\omega_kT} ~,
\label{orbit}
\end{equation}
and action, here $u_k=u_k(T)$,
\begin{equation}
S_{\text{lin}}=\nu\int\frac{dk}{2\pi}
\frac{|u_k-u'_k\exp(-\omega_kT)|^2}{1-\exp(-2\omega_kT)} ~.
\label{linac}
\end{equation}
We note that in the limit $T\rightarrow\infty$ the action
$S_{\text{lin}}(u,T)\rightarrow\nu\int(dk/2\pi)|u_k|^2$ in accordance with the
stationary distribution in Eq. (\ref{dis}) expanded in wave number space. 
Since at long times 
$u_k=\exp(\omega_kT)u'_k$ we also find the correction
\begin{equation}
S_{\text{diff}} = -\nu\int\frac{dk}{2\pi}|u_k|^2\exp(-2\omega_kT) ~.
\label{lincorr}
\end{equation}
From Eqs. (\ref{linac}) and (\ref{lincorr}) we observe the simple scaling
property, $u\rightarrow\mu u$, 
$S_{\text{lin}}\rightarrow\mu^2S_{\text{lin}}$, and
$S_{\text{diff}}\rightarrow\mu^2S_{\text{diff}}$, 
i.e., scaling the slope or height
field with a factor $\mu$ the action scales with $\mu^2$. This behavior
is compatible with the equations of motion (\ref{b11}) and (\ref{b12})
provided we scale $p\rightarrow\mu p$.

As regards the time dependence of 
$S_{\text{lin}}$ and $S_{\text{diff}}$ we identify the
crossover time
\begin{equation}
T^{\text{diff}}_{\text{co}}\sim\frac{1}{\nu k^2} ~,
\label{cot1}
\end{equation}
depending on the wave number $k$. In the thermodynamic limit
$L\rightarrow\infty$ the wave number $k$ has a 
continuous range and the crossover
time diverges in the infrared limit $k\rightarrow 0$.  Consequently, 
we do not have a separation of time scales.

Since the saturation width of an interface is a finite size effect
time scale separation only occurs for a finite system. In the present
linear case this is associated with the {\em quantization} of 
the wave number $k\sim 1/L$; note that the $n=0$ mode  
is related to the global conservation of slope , i.e.,
$\int dx~u$ is a constant of motion, and we have
\begin{equation}
T^{\text{diff}}_{\text{co}}\sim\frac{L^2}{\nu} ~.
\label{cot2}
\end{equation}
From the general discussion of a growing interface, $T_{\text{co}}\propto L^z$,
where $z$ is the dynamic scaling exponent, and we readily obtain $z=2$
in accordance with the diffusive mode contribution with dispersion
$\omega_k=\nu k^2$. 
For $T\ll T^{\text{diff}}_{\text{co}}$ the diffusive modes contribute
to the time-dependent distribution, whereas for 
$T\gg T^{\text{diff}}_{\text{co}}$ we cross over to the stationary
regime, $S_{\text{diff}}(u,T)\rightarrow 0$,  
and we approach the stationary distribution
(\ref{dis}) determined by $S_{\text{st}}(u)$.
\subsubsection{The stationary distribution}
In the one dimensional case the stationary distribution is known 
\cite{Huse85} and has
the symmetric Gaussian form given by Eq. (\ref{dis}). 
This follows directly from the stationary Fokker-Planck equation
(\ref{fp3}) where the $\lambda$-dependent term for a Gaussian
distribution becomes a total derivative and thus yields a vanishing
contribution; an argument which only holds in one dimension.
The slope field
$u(x)$ is thus uncorrelated beyond a finite correlation length $\xi$ which
is zero for the Burgers equation and microscopic for lattice models
falling in the same universality class. The height $h(x)=\int dx u$
performs random walk yielding according to Eq. (\ref{scal}) the roughness
exponent $\zeta = 1/2$.

Within the present canonical phase space formulation the 
stationary distribution follows
immediately from the structure $p=2\nu u$ of the stationary submanifold.
As in the harmonic oscillator case in Sec. III, the diffusive 
modes imply that $E_{\text B}\rightarrow 0$ for $T\rightarrow\infty$.
Thus inserting $p=2\nu u$ in the action (\ref{ac6}) in the 
one dimensional  case and
performing the time integration, we obtain in the limit $T\rightarrow\infty$
\begin{equation}
S_{\text{st}}(u) = \int_0^\infty dxdt~p\frac{\partial u}{\partial t} = 
\nu\int dx~u(x)^2
\end{equation}
and by insertion in Eq. (\ref{dis6}) the distribution (\ref{dis}).

Whereas the effective fluctuation-dissipation theorem valid in
one dimension implies that the stationary distribution is Gaussian and
symmetric in the slope $u$ and in the height field $h$, measured relative
to the mean height $\langle h\rangle$, the time-dependent distribution,
converging towards the stationary one, is expected to exhibit an asymmetric
shape corresponding to the predominance of peaks in $h$ in the growth
direction.  
\subsubsection{The long-time skew distribution}
The phase space approach also allows us to estimate the long-time corrections
to the stationary distribution. Following the reasoning in Sec. III
we replace for large $T$ the orbit {\em near the submanifold} $p=2\nu u$
with an orbit {\em on the submanifold}. Inserting $p=2\nu u$ in Eq. (\ref{b11})
the orbits on the stationary submanifold are governed by the noiseless
Burgers equation with $\nu$ replaced by - $\nu$,
\begin{equation}
\left(\frac{\partial u}{\partial t} - \lambda u\nabla\right)u = 
-\nu\nabla^2 u ~.
\label{nb}
\end{equation}
This equation is readily solved by means of the Cole-Hopf transformation
(\ref{ch}) with solution given by Eqs. (\ref{ch1}-\ref{ch3})
with -$\nu$ substituted for $\nu$. For the action
(\ref{ac6}) we then obtain $S_{\text {B}} = \nu\int dx[u^2 - u'^2]$ and for the 
time-dependent probability distribution 
\begin{equation}
P(u,T) \propto P_{\text{st}}(u)P_{\text{skew}}(u,T) ~,
\label{tdis}
\end{equation}
where the symmetric stationary distribution $P_{\text{st}}(u)$ is given by
Eq. (\ref{dis}) and the time-dependent skew correction by
\begin{equation}
P_{\text{skew}}(u,T)=\exp{\left[\frac{\nu}{\Delta}\int dx~u'(x)^2\right]} ~,
\label{skew}
\end{equation}
with $u'=\nabla h'$ and $u=\nabla h$ related by the
Cole-Hopf transformation
\begin{equation}
\exp{\left[-\frac{\lambda}{2\nu}h'(x)\right]} =
\int dx'~G(x-x',T)\exp{\left[-\frac{\lambda}{2\nu}h(x')\right]} ~.
\label{cht}
\end{equation}
Note that since $\int dx'G = 1$ the correction $u'^2$ vanishes in the limit
$T\rightarrow\infty$ and $P_{\text{skew}}\rightarrow 1$. 

In order to examine the skewness of the distribution it is convenient
to eliminate the stationary component by forming the ratio
\begin{equation}
\frac{P(u,T)}{P(-u,T)}=
\exp\left[\frac{\nu}{\Delta}\int dx(u'^2_+-u'^2_-)\right] ~,
\label{ratio}
\end{equation}
where according to the Cole-Hopf transformation (\ref{cht})
\begin{equation}
\exp{\left[-\frac{\lambda}{2\nu}h'_\pm\right]} =
\int dx'~G(x-x,T)\exp{\left[\mp\frac{\lambda}{2\nu}h\right]} ~.
\label{cht2}
\end{equation}
Inserting the Green's function (\ref{ch2}), 
$G(x,T)=[4\pi\nu T]^{-1/2}\exp[-x^2/4\nu T]$, we consider first a few 
simple cases.

For a constant slope $u=u_0$, i.e., $h=u_0x-h_0$, we obtain, performing
the Gaussian integration, $h'_\pm = \pm(u_0x+h_0)-\lambda Tu_0^2/2$. 
We note that the growth term $\lambda u\nabla u$ as expected gives rise
to a time-dependent term in $h'_\pm$. This term, however, is compensated
for by transforming to a co-moving frame as in the KPZ equation (\ref{kpz}).
The slope $u'$, however, is independent of $T$ and we obtain $u'_\pm=\pm u_0$,
yielding $S_{\text{B}} = 0$ and $P=\text{const.}$, 
i.e., no dynamics. This is consistent
with the fact that $u=u_0$ and $p=p_0$ trivially satisfy the field equations
(\ref{b11}) and (\ref{b12}) yielding $E_{\text{B}}=0$, and thus 
corresponds to
a stationary state, as also discussed in paper II.

Choosing a slope depending linearly on $x$, $u=2s_0x$, corresponding to
a parabolic height profile, $h=s_0x^2 + h_0$, we obtain 
$u'_\pm=\pm 2s_0x/(1\pm2\lambda Ts_0)$, yielding the skewness ratio
\begin{equation}
\frac{P(u,T)}{P(-u,T)}=
\exp\left[-\frac{32}{3}\frac{\nu\lambda}{\Delta}s_0^3
\frac{TL^3}{(1-(2\lambda Ts_0)^2)^2}\right] ~,
\label{ratio2}
\end{equation}
where we have introduced the size $L$ of the system. The expression
(\ref{ratio2}) only holds for $\lambda Ts_0<1$,
the important aspect is, however, the dependence on the
sign of $s_0$, i.e., the slope of the slope or the bias of the
height profile.  For $s_0>0$ corresponding to a parabolic shape of
$h$ with a minimum, i.e., a downward peak, $P(u,T)/P(-u,T)<1$, whereas
for $s_0<0$, yielding an upward peak in $h$, we have
$P(u,T)/P(-u,T)>1$. This behavior implies that the distribution is
skew at finite times and that the upward peaks in $h$ statistically
are more pronounced than the downward peaks, i.e., the distribution
is biased and changes asymmetrically towards the symmetric stationary 
distribution.

We can also gain some insight in the inviscid limit $\nu\rightarrow 0$
from a saddle point calculation along the lines of similar treatments
of the noiseless Burgers equation \cite{Kardar86,Medina89}. From
Eq. (\ref{cht}) we obtain, inserting the Green's function (\ref{ch2}),
\begin{equation}
\exp{[-\frac{\lambda}{2\nu}h'_\pm]} = 
\int dx'~[4\pi\nu T]^{-1/2}\exp{[-(1/2\nu)\phi_\pm]} ~,
\end{equation}
where $\phi_\pm(x,x') = (x-x')^2/2T \pm\lambda h(x')$. In the limit
$\nu\rightarrow 0$ the integral is dominated by the
local minima  given by $d\phi_\pm(x,x')/dx' = 0$ and 
the condition $d^2\phi_\pm(x,x')/dx'^2>0$. The solutions $x'_\pm$ are
thus determined by the implicit equation $x'_\pm-x=\mp\lambda Tu(x'_\pm)$
together with the conditions $\pm (du/dx')_{x'=x'_\pm}>-1/\lambda T$, and we
obtain the ratio
\begin{equation}
\frac{P(u,T)}{P(-u,T)} =
\exp{\left[\frac{\nu}{\Delta}\int dx\frac{(x-x'_+)^2-(x-x'_-)^2}
{(\lambda T)^2}\right]} ~,
\label{ratio3}
\end{equation}
which can be used in order to analyze the skewness of various
profiles $u(x)$. Note that $P(u,T)/P(-u,T)\rightarrow 1$ in the limit
$T\rightarrow\infty$, corresponding to vanishing skewness.
In Fig. 6 we have depicted the saddle point construction.
\begin{figure}
\begin{picture}(100,150)
\put(0.0,-190.0)
{
\centerline
{
\epsfxsize=13cm
\epsfbox{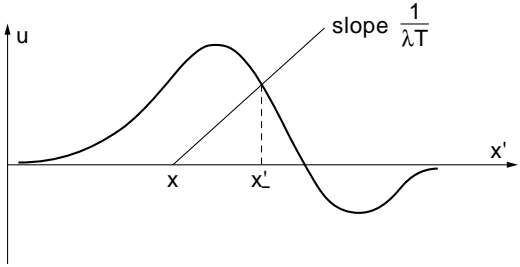}
}
}
\end{picture}
\caption{
Here we show the saddle point construction valid in the inviscid limit
$\nu\rightarrow 0$. The saddle point condition $x'_--x=\lambda Tu(x'_-)$
determines $x'_-$ as the intersection between the line with slope
$1/\lambda T$ and the slope profile $u(x')$. The intersection of the
line with the $x'$ axis determines $x$.
}
\end{figure}

More insight into the dynamics underlying the skew long-time distribution
is gained by expanding (\ref{cht}) in the nonlinear coupling.
Choosing a compact notation, $u_x=u(x)$, $h_x=h(x)$, $G_{xx'}(T)=G(x-x',T)$,
and $\delta_{xx'}=\delta(x-x')$, we obtain to leading order in
$\lambda$
\begin{eqnarray}
u'_x &&= \int dx'~G_{xx'}(T)u_{x'}
\nonumber
\\
&&-(\lambda/\nu)\int dx'
~G_{xx'}(T)u_{x'}h_{x'}
\nonumber
\\
&&+(\lambda/\nu)\int dx'dx''~G_{xx'}(T)G_{xx''}(T)u_{x'}h_{x''} ~.
\label{exp}
\end{eqnarray}
Correspondingly, $P_{\text{skew}}$ factorizes in a component independent of
$\lambda$ and a component depending on $\lambda$, 
\begin{eqnarray}
&&P_{\text{skew}}=P_0P_\lambda ~,
\label{fac}
\\
&&P_0=\exp{\left[\frac{\nu}{\Delta}
\int dxdx'~G_{xx'}(2T)u_xu_{x'}\right]} ~,
\label{lin}
\\
&&P_{\lambda}=
\exp{\left[-\frac{2\lambda}{\Delta}\int dxdx'dx''~F_{xx',x'x''}(T)
u_xu_{x'}h_{x''}\right]} ~.
\nonumber
\\
\label{skew2}
\end{eqnarray}
Here the
kernel $F$ is given by
\begin{eqnarray}
&&F_{xx',x'x''}(T) = 
\nonumber
\\
&&G_{xx'}(2T)\delta_{x'x''}
- \int dy~G_{yx}(T)G_{yx'}(T)G_{yx''}(T) ~.
\nonumber
\\
\label{ker}
\end{eqnarray}
For vanishing $\lambda$ only $P_0$ contributes.
In wave number space we thus obtain,
introducing $u_k = \int dx \exp{(ikx)}u(x)$ and noting from 
Eq. (\ref{ch2}) that $G_k(t) = \exp{[-\nu k^2t]}$, the distribution
\begin{equation}
P(u_k,T)\propto\exp{\left[-\frac{\nu}{\Delta}\int\frac{dk}{2\pi}
~|u_k|^2[1-\exp{(-2\nu k^2T)}]\right]} ~.
\\
\label{skew3}
\end{equation}
This result is completely equivalent to Eq. (\ref{ac5}) for the 
damped oscillator discussed in Sec. III. For $\lambda = 0$, corresponding
to the Edwards Wilkinson case, only  gapless diffusive modes
$u_k\propto\exp{[-\nu k^2t]}$ contribute to the time-dependent
distribution; we also note that the distribution remains symmetric.

To first order in $\lambda$ the diffusive modes interact and we obtain a
correction to $P_{\text{skew}}$ given by $P_\lambda$ in Eq. (\ref{skew2}). 
This contribution
is odd in $u$ and characterized by the 
kernel $F$. In other words, the scattering of the diffusive modes
on one another due to the term $\lambda u\nabla u$ in the Burgers
equation (\ref{nb}) yields a skew distribution in $u$. In the limit
$T\rightarrow\infty$ this term vanishes and we obtain the symmetric
stationary distribution. 
Correspondingly, in wave number space we have to order $\lambda$
\begin{eqnarray}
&&P_{\lambda}(u_k,T)\propto\left[-\frac{2\lambda}{\Delta}\int
\frac{dk}{2\pi}\frac{dk'}{2\pi}F_{k,k'}(T)u_ku_{-k-k'}h_{k'}\right] ~,
\nonumber
\\
\label{skew4}
\\
&&F_{k,k'}(T) = G_k(2T) - G_k(T)G_{k+k'}(T)G_{k'}(T) ~,
\label{skew4F}
\end{eqnarray} 
showing the interaction between the various $k$-modes (the cascade).
\subsubsection{The short-time skew distribution}
At shorter times, i.e., $T\leq T^{\text{diff}}_{\text{co}}$, the approximation
of replacing the orbit near the zero-energy submanifold with an
orbit on the submanifold ceases to be valid  and we have to consider the
equations of motions (\ref{b11}) and (\ref{b12}) in more detail in
order to identify the contribution to $S_{\text{sol}}$.

Although the noiseless Burgers equation is exactly soluble by means
of the Cole-Hopf transformation, the equations of motion (\ref{b11})
and (\ref{b12}) describing the noisy case are presumably
not exactly integrable. They do, however, admit special permanent
profile or solitary wave solutions with superposed linear diffusive
modes. It moreover follows from the path integral formulation in paper II
that an arbitrary interface profile can be represented by a dilute
gas of solitons, at least in the inviscid limit for small $\nu$, where
we can neglect soliton overlap contributions.

In the form given by Eqs. (\ref{con1}) and (\ref{con2}) these equations
were discussed in detail in paper II where we identified the 
{\em elementary excitations}.
The spectrum consists of {\em right hand} and {\em left hand} nonlinear soliton
modes with superposed linear diffusive modes. The localized soliton modes
have a finite energy and thus correspond to the nearby phase space orbits
approaching the zero-energy manifold. At long times the soliton energy
must go to zero 
and the remaining superposed diffusive modes 
determine $P_{\text{skew}}$ as discussed above. Note that the nonlinear soliton
mode can be regarded as a bound state of diffusive modes; this follows
from the stability analysis in paper I and is a consequence of Levinson's
theorem.
In paper II we performed a shift transformation of the noise variable,
$p\rightarrow\nu(u-\varphi)$, in order to express the Hamiltonian
density ${\cal H}_{\text{B}}$ in Eq. (\ref{hamb1}) in a canonical form with a
harmonic component describing the linear case, yielding the field equations
(\ref{con1}) and (\ref{con2}). In the present context we summarize the 
soliton dynamics in terms of $p$ and $u$ in accordance with the present
interpretation of the transition to the stationary state. The {\em right}
and {\em left hand} solitons then play a different role in the weight of the
interface morphology.
In the static limit the soliton modes have the form
\begin{equation}
u(x)=\pm u_0\tanh{\left[\frac{\lambda u_0}{2\nu}(x-x_0)\right]} ~,
\label{sol}
\end{equation}
with amplitude $u_0$ and position $x_0$. Using the Galilean invariance of
the field equations (\ref{b11}) and (\ref{b12}), i.e., observing that
the operator
$\partial/\partial t - \lambda u\nabla$ is invariant under the transformation
$x\rightarrow x-\lambda \tilde{u}t$, $u\rightarrow u+ \tilde{u}$,
propagating solitons with boundary values $u\rightarrow u_\pm$
for $x\rightarrow\pm L$, $L$ is the size of the system, are obtained
by boosting the static solitons in Eq. (\ref{sol}). Moreover, the
propagation velocity $v$ is given by $u_+$ and $u_-$ according to the 
soliton condition
\begin{equation}
u_+ + u_- = -\frac{2v}{\lambda} ~,
\label{solcon}
\end{equation}
which thus determines  the kinetics and matching conditions for a multi-soliton
configuration describing a growing interface.

The {\em right hand} soliton corresponds to $p=0$ and is according to
Eq. (\ref{b11}) a solution of the noiseless Burgers equation (\ref{bur0}).
Dynamical attributes are a feature of the noisy case and this soliton
thus
carries vanishing energy, $E_{\text{B}} = \int dx {\cal H}_{\text{B}} = 0$,
vanishing momentum, $\Pi_{\text{B}} = \int dxu\nabla p = 0$, 
and vanishing action
$S_{\text{B}} =0$, and corresponds to an orbit on the transient zero-energy
manifold. We note that a single {\em right hand} soliton or a multi-soliton
solution cannot satisfy the boundary condition of vanishing slope.

The {\em left hand} soliton moves on the submanifold $p=2\nu u$ and satisfies
according to Eq. (\ref{b11}) the noiseless Burgers equation (\ref{bur0})
with $\nu$ replaced by $-\nu$. It carries energy, momentum, and action 
given by
\begin{eqnarray}
E_{\text{B}}   && = \frac{2}{3}\nu\lambda(u^3_+ - u^3_-) ~,
\label{eb}
\\
\Pi_{\text{B}} && = \nu(u^2_+ - u^2_-) ~,
\label{mb}
\\
S_{\text{B}}   && = \frac{1}{6}\nu\lambda T|u_+-u_-|^3 ~.
\label{ab}
\end{eqnarray}
Since $u_+<u_-$ for a {\em right hand} soliton $E_{\text{B}}$ is negative and 
$\Pi_{\text{B}}$
has according to Eq. (\ref{solcon}) the same sign as the velocity $v$.
The action $S_{\text {B}}$ is positive and Galilean 
invariant. We also
note that although the soliton is confined to the submanifold $p=2\nu u$
the energy is nonvanishing. This is associated with the nonequal
boundary values $u_+$ and $u_-$ and also follows directly from
Eq. (\ref{hamb1}), where inserting $p=2\nu u$ we obtain a total derivative.
Hence, 
$E_{\text{B}} = \int dx{\cal H}_{\text{B}} = 
(2\nu\lambda/3)(u^3_+ - u^3_-)$, 
where $u_+$ and $u_-$ are
the boundary values.

\begin{figure}
\begin{picture}(100,170)
\put(0.0,-190.0)
{
\centerline
{
\epsfxsize=13cm
\epsfbox{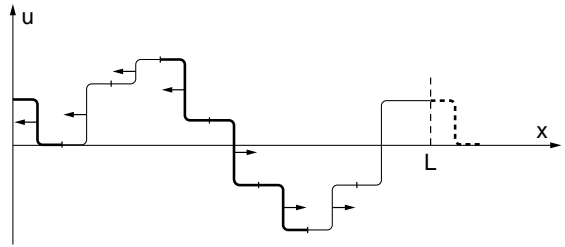}
}
}
\end{picture}
\caption{
We depict the slope soliton morphology for a growing interface
in a system of size $L$. In accordance with a growth situation
we have imposed periodic boundary conditions. The {\em left hand}
solitons carrying energy, momentum, and action are indicated 
by solid lines.
}
\end{figure}

As discussed in paper II the morphology of a growing interface is 
determined by matching a set of {\em right hand} and 
{\em left hand} solitons
according to the soliton condition (\ref{solcon}).
The {\em right} and {\em left hand}
solitons are exact solutions of the damped and undamped (-$\nu$) noiseless
Burgers equations, respectively. 
Their stability is associated with the
nonequality of the
boundary values, $u_+\neq u_-$, corresponding to a nonvanishing slope
current at the boundaries. In a multi-soliton configuration with vanishing 
boundary conditions current thus flows between the solitons.
The current is {\em generated} by the {\em left hand} solitons
and {\em dissipated} by the {\em right hand} solitons - 
this is another view of the cascade driving the noisy Burgers equation.
The morphology is depicted in Fig. 7.

The probability of a soliton morphology is determined by 
$S_{\text{sol}}(u,T)$.
Assuming that $S_{\text{sol}}=\lambda F(u,T)$ a scaling argument similar to
linear case, $u\rightarrow\mu u$, $p\rightarrow \mu p$, 
$\lambda\rightarrow\mu^{-1}\lambda$, and 
$S_{\text{sol}}\rightarrow \mu^2S_{\text{sol}}$,
following from the general form of the equations of motion (\ref{b11})
and (\ref{b12}) and the action (\ref{ac6}) implies that $F\rightarrow\mu^3F$.
This is consistent with the expression (\ref{ab}) and we obtain
\begin{equation}
S_{\text{sol}}(u,T) = \frac{1}{6}\nu\lambda T\sum_{\text{lhs}}|u_+-u_-|^3 ~,
\label{acsol}
\end{equation}
where the summation is only over contributing {\em left hand} solitons (lhs).

Beyond this point our discussion becomes necessarily more qualitative
and heuristic since we dont possess a complete solution of the coupled
field equations. The non-integrability and the constraint imposed by the 
soliton condition (\ref{solcon}) imply that we only have available 
a dilute gas of {\em right hand} and {\em left hand} solitons.
First we notice that for an infinite system the soliton velocity $v$
given by Eq. (\ref{solcon}) ranges freely, implying that the
{\em center of mass} of the soliton amplitude 
$u_{\text{cm}} = (u_++u_-)/2$ also
can take arbitrary values. Since $T$ enters as a prefactor in
Eq. (\ref{acsol}) $S_{\text{sol}}$ grows for $T\rightarrow\infty$,
i.e., $P\propto\exp[-S_{\text{sol}}/\Delta]\rightarrow 0$,
and we are unable to identify a solitonic crossover time.
This is consistent with the general discussion of a growing interface
\cite{Barabasi95}. Generally, the crossover time $T_{\text{co}}\propto L^z$,
where $z$ is the dynamic exponent. As in the linear case for
$\lambda=0$ discussed above the saturation width for a growing interface
is a finite size effect and the transient growth does not saturate
to stationary growth for an infinite system. Whereas the situation was
easy to analyze in the linear case where we can identify the independent
modes,  and where the system size $L$ is replaced by the inverse wave
number $1/k$, i.e., the thermodynamic limit is probed in the infrared
limit $k\rightarrow 0$, the situation is more subtle in the nonlinear
soliton case since we dont have a normal mode structure but only
approximate {\em elementary excitations}.

On the other hand, for a finite size system, imposing for example periodic
boundary conditions as indicated in Fig. 7 in order to ensure a 
growing interface in $h$ as the
{\em slope} solitons revolve, the velocity $v$ becomes endowed with
a scale and is 
{\em quantized} in units of $L/T$. Notice here the important difference 
between the diffusive case and the solitonic case. In the diffusive
case the {\em excitations} are not propagating but are linear combinations
of growing and decaying modes as discussed in Sec. III, and the 
system size $L$ only enters in
the {\em quantization} of the wave number $k\propto1/L$, yielding the
crossover time $T^{\text{diff}}_{\text{co}}\propto L^2$. 
In the solitonic case the localized modes are
propagating giving rise to genuine nonequilibrium growth. The system
size $L$ then  enters together with the time $T$ in setting 
a scale for the velocity $v$.

A simple estimate, replacing the soliton amplitude 
$u=u_+-u_-$ in the general expression
(\ref{acsol}) by $u_{\text{cm}}=(u_++u_-)/2=-v/\lambda$ 
from Eq. (\ref{solcon})
and moreover $v$ by $L/T$ we obtain 
$S_{\text{sol}}\propto\text{const.}~\nu L^3/\lambda^2T^2$ which inserted in
$P\propto\exp[-S_{\text{sol}}/\Delta]$ yields the soliton crossover time
\begin{equation}
T_{\text{co}}^{\text{sol}}
\propto\left(\frac{\nu}{\Delta}\right)\frac{1}{\lambda}L^{3/2} ~,
\label{scot}
\end{equation}
and we infer the dynamic exponent $z=3/2$.
In the transient short-time regime 
$T\ll T_{\text{co}}^{\text{sol}}$ the soliton
configurations contribute to $P$; 
in the long-time regime $T\gg T_{\text{co}}^{\text{sol}}$
the soliton contribution vanishes and only the diffusive modes and
their interactions contribute to $P$. 
We also notice 
that the expression (\ref{scot}) is consistent with
the dimensionless argument $\lambda(\Delta/\nu)^{1/2}t/x^{3/2}$ in the
scaling function for the slope correlation function discussed in paper II
and in \cite{Hwa91,Frey96,Janssen86}.
In Fig. 8 we have depicted the crossover regimes.
\begin{figure}
\begin{picture}(100,170)
\put(0.0,-190.0)
{
\centerline
{
\epsfxsize=13cm
\epsfbox{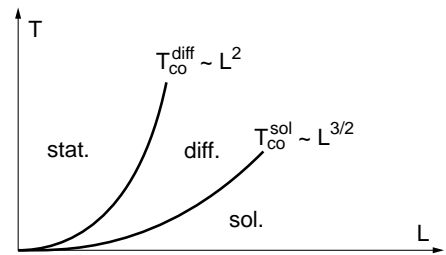}
}
}
\end{picture}
\caption{
Here we depict the crossover time as a function of the system size $L$.
In the early time regime for $T\ll T_{\text{co}}^{\text{sol}}$ the
distribution is dominated by soliton contributions. In the intermediate
time regime for 
$T_{\text{co}}^{\text{diff}}\gg T\gg  T_{\text{co}}^{\text{sol}}$
the soliton contributions become suppressed leaving the diffusive
mode contributions. Finally, for $T\gg T_{\text{co}}^{\text{diff}}$
the diffusive modes also die out and we approach the stationary distribution.
}
\end{figure}

The short-time probability distribution $P(h,T)$ for
$T\ll T_{\text{co}}^{\text{sol}}$ has been discussed within
the directed polymer approach \cite{Halpin95,Kardar87a}. Based
upon a replica scaling analysis \cite{Zhang90d} one finds for positive
$h$ (measured relative to the growing mean height) the heuristic
expression $P(h,T)\propto\exp[-(h/T^{1/3})^{\eta}]$, where
$\eta = 3/2$; for $h<0$ based on numerical results, $\eta\sim 2.5$.
Recent exact results for the asymmetric exclusion model
which falls in the same universality class as the Burgers equation, see
ref. \cite{Derrida98}, where other references also can be found,
also seems to have bearings on the height distribution.
Using the Bethe ansatz method a skew distribution, characterized
by the exponents $3/2$ and $5/2$, have been found for the large
deviation function of the time averaged current.

Within the present soliton approach we can derive a qualitative expression
for the early time height distribution by noting that 
$|u_+-u_-|^3\sim(uL)^{3/2}(T\lambda)^{-3/2}\sim
h^{3/2}(T\lambda)^{-3/2}$. Inserting this result in the general
expression (\ref{acsol}) we obtain
\begin{equation}
P_{\text{sol}}(h,T)\propto
\exp
\left[-\frac{\nu}{\Delta}\left[\frac{1}{\lambda T}\right]^{1/2}h^{3/2}\right] ~,
\end{equation}
in accordance with the directed polymer-replica based result and related
to the exact results for the asymmetric exclusion model.
The skewness of the distribution in $h$ must then arise from the bias in the
statistical weight $\exp[-S/\Delta]$ assigned to the {\em left}
and {\em right hand} solitons giving rise to a predominance of
{\em right hand} solitons ($S=0$), corresponding to relative forward
growth. 
Unfortunately, our present understanding of the soliton approach
and the inaccessibility of a more detailed multi-soliton solution
do not allow a more detailed analysis.

\subsection{The general case}
In the general case for $d>1$ the slope and noise fields $u_n$
and $p_n$ have longitudinal vector character and are governed by
Eqs. (\ref{b1}) and (\ref{b2}) determining an orbit 
$(u_n(x_p,t),p_n(x_p,t))$ in $p_nu_n$ phase space. At long times
the orbit must pass close to the stationary saddle point
$(u_n,p_n)=(0,0)$ in order to induce Markovian behavior and then 
progress onto the stationary zero-energy manifold 
$E_{\text{B}}=\int d^dx~{\cal H}_{\text{B}}=0$ with energy density given
by Eq. (\ref{ham4}), yielding the weak noise distribution
$P\propto\exp[-S_{\text{B}}/\Delta]$ with $S_{\text{B}}$ given by
Eq. (\ref{ac6}).

In $d=1$, as discussed in Sec. A, the orbit on the stationary
manifold is attracted to the submanifold $p=2\nu u$, yielding the
symmetric stationary distribution (\ref{dis}) with time-dependent
skew corrections. 
For $d>1$ this behavior is only encountered in
the linear Edwards-Wilkinson case for $\lambda = 0$; note also
the general discussion in Sec. II. The attraction to the submanifold
$p_n=2\nu u_n$ is associated with the underlying fluctuation-dissipation
theorem and gives rise to the stationary Gaussian distribution
$P_{\text{st}}\propto\exp[-(\nu/\Delta)\int d^dx~u_n(x)^2]$;
the corresponding free energy is $F=(1/2)\int d^dx~u_n(x)^2$.
This distribution yields the roughness exponent $\zeta = (2-d)/2$. 
The dynamic exponent
$z=2$, corresponding to the diffusive mode contribution; note that
the Galilean invariance is not operative in the linear case and
that we consequently dont have the scaling law constraint
$\zeta + z =2$.

In the nonlinear case $\lambda\neq 0$ for $d>2$ the long-time orbit
emerging from the vicinity of the stationary point 
$(u_n,p_n)=(0,0)$ diverges for larger $u_n$ from the submanifold
$p_n=2\nu u_n$ which constitutes a sort of {\em tangent plane}
to the zero-energy surface at the stationary point. In the limit
$\lambda\rightarrow 0$ the energy surface then collapses to the
tangent plane. More detailed, in the stationary limit the distribution
is given by 
\begin{equation}
P_{\text{st}}\propto\exp\left[-\frac{1}{\Delta}\int_0^\infty
d^dxdt~p_n\frac{\partial u_n}{\partial t}\right] ~.
\label{stadis3}
\end{equation}
Noting that $p_n$ is {\em slaved} to $u_n$ on the orbit we obtain 
in general (to leading order in $\nabla_n$)
\begin{equation}
p_n=u_nF_d(u^2) + u_p\nabla_nu_pG_d(u^2) ~,
\label{prel}
\end{equation}
where the scalar functions $F_d$ and $G_d$ depend on the invariant
$u^2=u_nu_n$ and parametrically on the dimension $d$. In one
dimension we have $F_1=2\nu$ and $G_1=0$. For $d>1$ to leading order
in $\nabla_n$ the distribution $P_{\text{st}}$ is even in $u_n$. Note,
however, that for $G_d\neq 0$ there is a skew correction to $P_{\text{st}}$.
Assuming that this analysis is valid, the determination of $F_d$ and $G_d$
is then given by the solution of the equations of motion (\ref{b1}) and 
(\ref{b2}) on the stationary zero-energy manifold.

In the nonlinear case for $d>2$ the rough phase governed by the
strong coupling fixed point only appears for a renormalized coupling
strength $\tilde{\lambda}=(\Delta\lambda)^2/\nu^3$ exceeding a finite
threshold value $\tilde{\lambda}_c$ \cite{Imbrie88,Cook89,Evans92}
and recent work \cite{Laessig98a,Chin98} moreover indicates
that unlike the case in $d=1$ even the
stationary probability distribution exhibits skewness, i.e.,
$P_{\text{st}}(u)\neq P_{\text{st}}(-u)$.
Such a behavior does not seem compatible with the analysis above and
indicates that the present weak noise approach
only applies  to the weak coupling phase for $d>2$. Presumably,
the strong coupling phase for $\tilde{\lambda}>\tilde{\lambda}_c$
is only accessed beyond a critical noise strength $\Delta_c$,
$\tilde{\lambda_c}=(\Delta_c\lambda)^2/\nu^3$. These issues and the fact that
the strong coupling fixed point (fortuitously) can be analysed in
$d=1$ in a nonperturbative weak noise approximation remains not
very well understood and call for further investigations.
\subsection{Summary and conclusion}
In the present paper we have advanced a general weak noise canonical
phase space approach to stochastic systems governed by a Langevin equation
driven by additive white noise. Reformulating the associated Fokker-Planck
equation in the nonperturbative weak noise limit in terms of a
Hamilton-Jacobi equation we have discussed the time-dependent and
stationary probability distributions from a canonical phase space point of 
view. The issue of solving the stochastic Langevin equation or the
associated Fokker-Planck equation is thus replaced by solving 
coupled Hamilton equations of motion determining the orbits in phase
space. The stochastic nature of the underlying problem is reflected 
in a peculiar topology of the energy surfaces different from the one
encountered in ordinary dynamical problems.  The Markovian behavior
thus corresponds to the existence of a stationary hyperbolic saddle point
which controls the behavior of the orbits in the long time limit.

We have in particular applied the canonical phase space approach
to the noisy Burgers equation describing a growing interface in
one dimension. We have recovered the well-known stationary distribution
and derived expressions for the time-dependent distribution, at long
times governed by linear diffusive modes and their interaction and
at shorter time by nonlinear soliton excitations. In higher dimensions
where the noisy Burgers equation predicts a kinetic phase transition
to a strong coupling phase the canonical phase space approach only
seems to access the weak coupling phase.

\acknowledgements
{
Discussions with M. Kosterlitz, B. Derrida, T. Hwa, J. Hertz,
P. Cvitanovi\'{c}, T. Bohr, M. H. Jensen, 
J.-P. Bouchaud, J. Krug, M. Alava, K. B. Lauritsen, K. M\o lmer  and A. Svane
are gratefully acknowledged.
}

\end{multicols}
\end{document}